\documentclass[prd,floatfix,showpacs,eqsecnum,superscriptaddress]
{revtex4}
\usepackage{epsfig}
\usepackage{verbatim, amsmath,amssymb}

\newcommand{\comments}[1]{}
\newcommand{\eq}[1]{Eq.~(\ref{#1})}
\newcommand{\fig}[1]{Fig.~{\ref{#1}}}

\newcommand{\be}{\begin{equation}}
\newcommand{\ee}{\end{equation}}
\newcommand{\bea}{\begin{eqnarray}}
\newcommand{\eea}{\end{eqnarray}}
\newcommand{\ben}{\begin{eqnarray*}}
\newcommand{\een}{\end{eqnarray*}}

\newcommand{\DS}{Dyson-Schwinger }
\newcommand{\BS}{Bethe-Salpeter }
\newcommand{\ST}{Slavnov-Taylor }
\newcommand{\YM}{Yang-Mills }

\newcommand{\w}{\omega}
\newcommand{\e}{\varepsilon}
\newcommand{\al}{\alpha}
\newcommand{\ba}{\beta}
\newcommand{\ga}{\gamma}
\newcommand{\G}{\Gamma}
\newcommand{\de}{\delta}

\newcommand{\si}{\sigma}

\newcommand{\ro}{\rho}
\newcommand{\la}{\lambda}

\newcommand{\ka}{\kappa}

\newcommand{\et}{\eta}

\newcommand{\pd}{\partial}

\renewcommand{\th}{\theta}
\newcommand{\cd}{{\cal D}}
\newcommand{\cs}{{\cal S}}
\renewcommand{\div}{\vec{\nabla}}

\newcommand{\s}[2]{{#1}\!\cdot\!{#2}}
\newcommand{\ov}[1]{\overline{#1}}
\newcommand{\dk}[1]
{\,\,\,\raisebox{-0.4ex}{\large $\bar{}$}\!\!d\,{#1}\,}

\newcommand{\ev}[1]{<\!\!{#1}\!\!>}

\begin{document}
\title{ Higher order heavy quark Green's functions in Coulomb gauge}
\author{C.~Popovici}
\affiliation{Institut f\"ur Theoretische Physik, Universit\"at
T\"ubingen, Auf der Morgenstelle 14, D-72076 T\"ubingen, Germany}
\affiliation {Centro de F\'{i}sica Computacional,
Departamento de F\'{i}sica, Universidade de Coimbra,
3004-516 Coimbra, Portugal}
\author{P.~Watson}
\author{H.~Reinhardt}
\affiliation{Institut f\"ur Theoretische Physik, Universit\"at
T\"ubingen, Auf der Morgenstelle 14, D-72076 T\"ubingen, Germany}
\begin{abstract}
The \DS equation for the 4-point quark Green's functions is studied.
In the limit of the heavy quark mass and with the truncation to
include only the dressed two point functions for the Yang-Mills
sector, we provide an exact solution for the 4-point quark Green's
functions, in both quark-antiquark and diquark channels, and show
that the corresponding poles relate to the bound state energy of the
heavy quark systems. Moreover, a natural separation between physical
and unphysical poles in the Green's functions emerges.
\end{abstract}
\maketitle

\section{Introduction}
\setcounter{equation}{0}

Since its foundation more than 30 years ago, one of the main goals of
Quantum Chromodynamics [QCD] is to understand the structure of hadrons
via the interactions of its elementary degrees of freedom, the quarks
and gluons. In order to approach this fundamental question, one of the
most appropriate starting points is to consider Coulomb gauge. In this
gauge, there is a natural connection to the would-be physical degrees
of freedom \cite{Zwanziger:1998ez}, and an appealing picture of
confinement exists: the so-called Gribov-Zwanziger scenario of
confinement \cite{Gribov:1977wm, Zwanziger:1995cv,Zwanziger:1998ez},
whereby the temporal component of the gluon propagator provides for a
long range confining force in the infrared, while the transversal
spatial component is infrared suppressed (and therefore does not
appear as an asymptotic state). In the last years, important progress
has been made in investigating the \YM sector of the theory in Coulomb
gauge: the variational method in the Hamiltonian approach
\cite{Feuchter:2004mk,Epple:2006hv,Epple:2007ut, Szczepaniak:2001rg},
the Lagrange-based (Dyson-Schwinger) functional formalism
\cite{Watson:2006yq,Watson:2007vc} and also lattice calculations
\cite{Burgio:2008jr,Quandt:2008zj} (see also
\cite{Nakagawa:2009zf,Cucchieri:2000gu,Langfeld:2004qs,
Cucchieri:2007uj}). On the lattice, the temporal gluon propagator
appears to be largely energy independent and behaves like $1/\vec q^4$
in the infrared, whereas the static spatial gluon propagator is found
to be vanishing in the infrared, in agreement with the Gribov formula
\cite{Burgio:2008jr}. In the framework of the functional formalism it
has been shown that the total color charge of the system is conserved
and vanishing, and the well-known energy divergence problem in Coulomb
gauge disappears \cite{Reinhardt:2008pr}. Moreover, within this
approach, the \ST identities \cite{Watson:2008fb} have been derived
and perturbative results have been provided
\cite{Watson:2007vc,Watson:2007mz}.  In the quark sector, perturbative
results have been presented \cite{Popovici:2008ty}, and in the heavy
mass limit the confining potential has been studied, both for two and
three quark systems \cite{Popovici:2010mb, Popovici:2010ph}.

In general, the underlying equation for the description of meson bound
states is the two-body homogeneous \BS equation.  In the
rainbow-ladder approximation, this equation has been successfully used
to describe the properties of light mesons (see, for example
\cite{Maris:1999nt, Alkofer:2002bp} and for a recent review
\cite{Fischer:2006ub}), where the driving mechanism is the chiral
symmetry breaking.  Beyond this approximation, models with dressed
vertex contributions \cite{Watson:2004kd,Williams:2009wx,
Bender:2002as,Bhagwat:2004hn,Matevosyan:2006bk,Bender:1996bb} and
unquenching effects \cite{Fischer:2005en,Watson:2004jq,
Fischer:2009jm} have been considered, and more sophisticated numerical
methods to solve both the homogeneous and the inhomogeneous \BS
equation have been recently developed \cite{Blank:2010bp}. In spite of
this success, an exact derivation of the meson or diquark bound state
energies via Green's functions techniques (i.e., using the \DS
equations as opposed to the \BS equation) has not been yet
reported. The difficulty stems from the fact that the (irreducible)
interaction kernel contains higher order vertex functions which in
general can not be calculated exactly. The connection between the
singularities of the Green's functions and the bound states dates back
to the original work of Gell-Mann and Low \cite{GellMann:1951rw}, who
gave a rigorous proof of the two-particle \BS equation. It is,
moreover, one of the longstanding problems in relativistic quantum
field theory, and in particular phenomenological studies of meson and
baryon states, to identify the \emph{physical} solutions of the poles
of the Green's functions and to separate them from possible unphysical
ones. For an analysis of the appearance of unphysical poles in QED
bound states, see for example Ref.~\cite{Ahlig:1998qf} and references
therein.

Based on the heavy mass expansion underlying the Heavy Quark Effective
Theory [HQET] \cite{Neubert:1993mb, Mannel:1992fx} and with the
truncation of the \YM sector to include only dressed two-point
functions, we study nonperturbatively quark-antiquark and diquark
states using Green's functions techniques. By means of functional
methods, we explicitly derive the \DS equation for the quark 4-point
Green's functions and give an exact, analytical solution. This will
enable us to verify that bound states are related to the occurrence of
physical poles in the Green's functions and hence we will be able to
provide a direct connection between the homogeneous \BS equation
(considered previously in Ref.~\cite{Popovici:2010mb}) and the
singularities of the Green's function, at least within the scheme
considered here. Moreover, within the employed approximation we will
show that the physical and unphysical poles naturally separate.

This paper is organized as follows. In Sec.~II we review the
functional formalism and derive the equations of motion, along with
the relevant higher order functional derivatives. We present the
explicit derivation of the \DS equations for the 4-point Green's
functions for quark-antiquark systems. In Sec.~III we give a brief
survey of the results obtained for heavy quark systems. After making
an expansion of the generating functional of Coulomb gauge QCD in
powers of $1/m$, where $m$ is the quark mass, and keeping only the
leading term, we present the main steps in the derivation of the
(anti)quark gap equation, and the corresponding temporal quark-gluon
vertices.  In Sec.~IV we consider both the 1-particle irreducible and
amputated connected 4-point quark-antiquark Green's functions in the
heavy mass limit. In this approximation, analytical solutions for the
Green's functions are provided and the singularity structure, in
particular the separation of the physical and unphysical states, is
discussed. Moreover, the correspondence with the bound state solution
emerging from the homogeneous \BS equation is emphasized. Sec.~V is
devoted to diquark systems. Again, the separation of the poles and the
connection to the \BS equation (including flavor structure) is
discussed. A summary and some concluding remarks are given in
Sec.~VI. In the Appendix, the explicit derivation of the \ST identity
for the quark-2 gluon vertex is presented.

\section{ \DS equations for 4-point functions}

We begin by briefly reviewing the functional formalism used to derive
the basic equations that will lead to the \DS equations and \ST
identities.  Following the notations and conventions from
\cite{Popovici:2008ty, Popovici:2010mb}, we work in Minkowski space,
with the metric $g_{\mu\nu}=\textrm{diag}(1, -\vec 1)$.  Roman letters
($i, j \dots$) denote spatial indices, greek letters ($\mu,\nu \dots$)
denote Lorentz indices and superscripts ($a, b, c \dots$) stand for
color indices in the adjoint representation.  Unless otherwise
specified, the Dirac flavor, spinor and (fundamental) color indices
are labelled with a common index ($\al, \ba \dots$).  Also,
configuration space coordinates may be denoted with subscript ($x,y,z
\dots$) when no confusion arises.  The Dirac $\ga$ matrices satisfy
$\{\ga^\mu,\ga^\nu\}=2g^{\mu\nu}$.

Let us consider the explicit quark contribution to the full QCD
generating functional \cite{Popovici:2008ty}:
\bea 
Z[\ov{\chi},\chi]&=&\int{\cal D}\Phi\exp{\left\{\imath \int d^4x
\ov{q}_\al(x)\left[\imath\ga^0D_0
+\imath\s{\vec{\ga}}{\vec{D}}-m\right]_{\al\ba}q_\ba(x) \right\}}
\nonumber\\&&
\times \exp{\left\{\imath\int
d^4x\left[\ov{\chi}_\al(x)q_\al(x)+
\ov{q}_\al(x)\chi_\al(x)\right]+\imath \cs_{YM}\right\}}.
\label{eq:genfunc} 
\eea
In the above, $\cs_{YM}$ represents the \YM part of the action and
${\cal D}\Phi$ generically denotes the functional integration measure
over all fields.  $q_\al$ denotes the full quark field, $\bar q_\al$
is the conjugate (or antiquark) field, and $\ov{\chi}_\al$, $\chi_\al$
are the corresponding sources. The temporal and spatial components of
the covariant derivative (in the fundamental color representation) are
given by
\bea
D_0&=&\partial_{0}-\imath gT^a\si^a(x),\nonumber\\
\vec{D}&=&\div+\imath gT^a\vec{A}^a(x),
\eea
where $\vec{A}$ and $\si$ refer to the spatial and temporal components
of the gluon field, respectively, and the source terms for $\vec A$
and $\si$ (with sources $\vec J$ and $\rho$, respectively) are
implicitly included in $\cs_{YM}$.  The structure constants of the
$SU(N_c)$ group are denoted with $f^{abc}$, and the Hermitian
generators $T^a$ satisfy $[T^a,T^b]=\imath f^{abc}T^c$ and are
normalized via $\mbox{Tr}(T^aT^b)=\de^{ab}/2$. Also note the Casimir
factor associated with the quark gap equation:
\be
C_F=\frac{N_c^2-1}{2N_c}.\label{eq:casimir1}
\ee

The field equation of motion is derived from the generating
functional, \eq{eq:genfunc}, and the observation that the integral of
a total derivative vanishes, up to possible boundary terms. We assume
that these terms do not contribute, although this is not obvious due
to the Gribov problem \cite{Gribov:1977wm} (see \cite{Watson:2006yq}
and references therein), and obtain:
\be 
\int\cd\Phi\frac{\delta}{\delta\imath \bar q_{\ga}(y)} \exp
{\left\{\imath \int d^4x \ov{q}_\al(x)\left[\imath\ga^0D_0
+\imath\s{\vec{\ga}}{\vec{D}}-m\right]_{\al\ba}q_\ba(x) +\imath\int
d^4x\left[\ov{\chi}_\al(x)q_\al(x)+
\ov{q}_\al(x)\chi_\al(x)\right]+\imath{\cal S}_{YM}\right\}} =0.
\label{eq:gen3} 
\ee

So far, the generating functional $Z[J]$ generates both connected and
disconnected Green's functions. The generating functional of connected
Green's functions is $W[J]$, where $W=\ln Z$.  We introduce a bracket
notation for the functional derivatives of $W$, such that for a
generic source denoted by $J_\al$ (in this case the index $\al$ refers
to all the attributes of both the quark and gluon fields, including
the type):
\be
\frac{\delta W}{\delta \imath J_{\alpha}}= \ev{\imath J_{\alpha}}.
\ee
The classical fields are (we use the same notation for the classical
and quantum fields, as is standard):
\be
\Phi_\al(x)=\frac{1}{Z}\int{\cal D}\Phi \Phi_\al(x)\exp{\left\{\imath{
\cal S}\right\}}=\frac{1}{Z}\frac{\de Z}{\de\imath J_\al(x)}.
\ee
The generating functional of the proper (one particle irreducible)
Green's function is the effective action $\G$ (function of the
classical fields) and is defined via the Legendre transform of $W[J]$:
\be
\G[\Phi]=W[J]-\imath J_\al\Phi_\al.
\label{eq:legtr}
\ee
In the above, we have used the common convention that the index $\al$
is either summed or integrated over as appropriate.  This gives
\be 
\ev{\imath J_{\al}}=\frac{\de W}{\de \imath J_\al}=\Phi_{\al}
\textrm{~~~and~~~} 
\ev{\imath \Phi_{\al}}=\frac{\de \G}{\de \imath\Phi_\al}=- J_{\al},
\label{eq:deffuncder} 
\ee
where the same bracket notation for derivatives of $\G$ with respect
to the classical fields is used -- there is no confusion between the
two sets of brackets since the derivatives with respect to the sources
and fields are never mixed.

We now present the quark field equation of motion in terms of proper
Green's functions, arising from \eq{eq:gen3} (and from which the gap
equation and the 4-point quark Green's functions will be derived):
\bea
\ev{\imath \bar q_{\al}(x)} &=& -\imath\left[\imath\gamma^{0}\pd_{0x}+
\imath\vec\gamma\cdot\vec\nabla_{x}-m\right]_{\al\ba}\imath
q_{\ba}(x)\nonumber\\ &&+\left[g
T^{c}\gamma^{0}\right]_{\al\ba}\left[\si^{c}(x)q_{\ba}(x)
+\ev{\imath\rho ^{c} (x)\imath\bar\chi_{\ba}(x)}\right] -\left[g
T^{c}\gamma^{k}\right]_{\al\ba}\left[A_{k}^{c}(x)q_{\ba}(x)
+\ev{\imath J_{k}^{c}(x)\imath\bar\chi_{\ba}(x)}\right].
\label{eq:qeom}
\eea
We retain for the moment the spatial part in the above equation,
although later on this will be truncated out within the employed
approximation scheme.

At this stage, it is useful to introduce some notation for multiple
functional derivatives with respect to fields and sources, which will
be later on used to derive the \DS equations. Consider the following
partial differentiation expressions arising from the Legendre
transform, \eq{eq:legtr} (we omit the configuration space arguments
for clarity):
\begin{subequations}
\bea
\frac{\de}{\de\imath \Phi_{\ba}}\ev{X(J)}&=& -\imath S[\ga]\ev{\imath
\Phi_{\ba}\imath \Phi_{\ga}}\ev{\imath J_{\ga} X
(J)}\label{eq:mfd10},\\
\frac{\de}{\de\imath J_{\ba}}\ev{Y(\Phi)}&=& \imath S[\ga]\ev{\imath
J_{\ba}\imath J_{\ga}}\ev{\imath \Phi_{\ga} Y(\Phi)},
\label{eq:mfd11}
\eea
\end{subequations}
where $S[\ga]=\pm1$ accounts for the fact that the fields may be
Grassmann-valued, i.e.  a minus sign appears when the index $\ga$
refers to the following contribution to the sum (over ``type'' $\ga$)
\be
\ev{\dots\imath q_{\ga}}\ev{\imath \bar\chi_{\ga}\dots},\,\,\,
\ev{\dots\imath \chi_{\ga}}\ev{\imath \bar q_{\ga}\dots}.
\ee
Combining \eq{eq:mfd10} and \eq{eq:mfd11} recursively, one easily
finds the following useful relationship:
\be
\frac{\de}{\de\imath \Phi_{\ba}}\ev{\imath J_{\de}\imath J_{\al}}=
-\eta_{\ba \de} S[\ga, \ka] \ev{\imath J_{\de}\imath J_{\ka}}
\ev{\imath \Phi_{\ka}\Phi_{\ba}\imath \Phi_{\ga}} \ev{\imath
J_{\ga}\imath J_{\al}},
\label{eq:mfd4}
\ee
where the factor $\eta_{\ba \de}=-1$ if fields/sources of the type
$\de, \ba$ anticommute. Of course, this type of relationship is
standard, but the notation allows one to follow the various multiple
derivatives and keep track of the signs efficiently.

The quark gap equation is derived by taking the functional derivative
of the quark field equation of motion (in configuration space),
\eq{eq:qeom}, with respect to $\imath q_{\ga}(z)$, and omitting the
terms which will eventually vanish when the sources are set to zero:
\bea
\label{eq:qq}
\ev{\imath\bar q_{\al}(x)\imath q_{\ga}(z)} &=&
\imath\left[\imath\gamma^{0}\pd_{0 x}
+\imath\vec\gamma\cdot\vec\nabla_{x}-m\right]_{\al\ga}\delta (x-z)
\nonumber\\
&&-\int d^4y\, \de(x-y) \left[ \G_{\bar
qq\si\al\ba}^{(0)a}\frac{\delta}{\delta\imath q_{\ga}(z)}
\ev{\imath\rho^{a}(y)\imath\bar\chi_{\ba}(x)} +\G_{\bar qq Aj
\al\ba}^{(0)a}\frac{\delta}{\delta\imath q_{\ga}(z)} \ev{\imath
J_{j}^{a}(y)\imath\bar\chi_{\ba}(x)} \right]
\eea
where the (configuration space) tree-level quark-gluon vertices,
obtained from the quark equation of motion \eq{eq:qeom}, are given by
\bea
\G_{\bar q q\si\al\ba}^{(0)a}&=&\left[gT^{a}\ga^{0}\right]_{\al\ba},
\label{eq:treelevelquarkvertex1}\\
\G_{\bar q qAj\al\ba}^{(0)a}&=&-\left[gT^{a}\ga^{j}\right]_{\al\ba},
\label{eq:treelevelquarkvertex2}
\eea
(omitting the trivial $\de$-function configuration space dependence).
We now use the formula \eq{eq:mfd4} to calculate the functional
derivatives appearing in the bracket and obtain (for simplicity, we
write the spatial arguments of the external fields/sources as
subscripts and omit those for the internal fields/sources that are
implicitly summed and integrated over):
\bea
\frac{\delta}{\delta\imath q_{\ga
z}}\ev{\imath\rho_{y}^{a}\imath\bar\chi_{\ba x}}&=& -S[\la,\ka]
\ev{\imath \rho_{y}^{a}\imath J_{\la}} \ev{\imath\Phi_{\la}\imath
q_{\ga z}\imath \Phi_{\ka}} \ev{\imath J_{\ka}\imath\bar\chi_{\ba x}},
\label{eq:rhoqq}\\
\frac{\delta}{\delta\imath q_{\ga z}}\ev{\imath
J_{jy}^{a}\imath\bar\chi_{\ba x}}&=& -S[\la,\ka] \ev{\imath
J_{jy}^{a}\imath J_{\la}} \ev{\imath \Phi_{\la}\imath q_{\ga z}\imath
\Phi_{\ka}} \ev{\imath J_{\ka}\imath\bar\chi_{\ba x}}.
\label{eq:jqq}
\eea
When sources are set to zero, the brackets on the right-hand side
correspond to connected and proper Green's functions. We can then
identify the field/source type ``$\la$" with the gluon (temporal for
the upper expression, spatial for the lower) and ``$\ka$'' with the
antiquark, and notice that in both cases $S[\la,\ka]=+1$.  Introducing
our conventions and notations for the Fourier transform, we have for a
general two-point function (connected or proper) which obeys
translational invariance:
\bea
\ev{\imath J_{\al y} \imath J_{\ba x}}&=& \int \dk{k} W_{\al\ba}(k)
e^{-\imath k\cdot (y-x)},
\label{eq:fourier1}\\
\ev{\imath \Phi_{\al y}\imath \Phi_{\ba x}}&=& \int \dk{k}
\G_{\al\ba}(k) e^{-\imath k\cdot (y-x)},
\label{eq:fourier2}
\eea
and for the three-point (proper vertex) function
\be
\ev{\imath \Phi_{\al y}\imath \Phi_{\ba x} \imath \Phi_{\ga z}}= \int
\dk{k_1}\dk{k_2}\dk{k_3} \G_{\al\ba\ga}(k_1,k_2,k_3) e^{-\imath
k_1\cdot y-\imath k_2\cdot x-\imath k_3\cdot z}
(2\pi)^{4}\de(k_1+k_2+k_3),
\label{eq:fourier3}
\ee
(similarly for higher $n$-point functions) where $\dk{k}=d^4
k/(2\pi)^{4}$. The propagator $W_{\al\ba}(k)$ and proper (1PI)
two-point function $\G_{\al\ba}(k)$ are related via the Legendre
transform. For the quark propagator, we have the standard relation
\be
W_{\bar qq\al\ga} (k) \G_{\bar qq\ga\ba} (k) =\de_{\al\ba}.
\ee
We now insert the expressions Eqs. (\ref{eq:rhoqq}), (\ref{eq:jqq})
into \eq{eq:qq}. After Fourier transforming to momentum space and with
the definitions Eqs. (\ref{eq:fourier1}, \ref{eq:fourier2},
\ref{eq:fourier3}) we obtain the quark \DS (or gap) equation:
\bea
\G_{\ov{q}q\al\ga}(k)&=\G_{\ov{q}q\al\ga}^{(0)}(k)+\int\dk{\w}&
\left\{ \G_{\ov{q}q\si\al\ba}^{(0)a}(k,-\w,\w-k)W_{\ov{q}q\ba\ka}
(\w)\G_{\ov{q}q\si\ka\ga}^{b}(\w,-k,k-\w)W_{\si\si}^{ab}(k-\w) \right.
\nonumber\\&&\left.
+\G_{\ov{q}qA\al\ba i}^{(0)a}(k,-\w,\w-k)W_{\ov{q}q\ba\ka}(\w)
\G_{\ov{q}qA\ka\ga j}^{b}(\w,-k,k-\w)W_{AAij}^{ab}(k-\w) \right\}
\label{eq:gap}.
\eea

Let us now derive the \DS equation for the one-particle irreducible
(1PI) 4-point function.  As an illustration of the functional
differentiation techniques, we present the explicit derivation of the
first term in this expression and notice that the rest of the terms
follow from an identical calculation (as explained below, the $\vec
A$-terms will be eliminated within our truncation scheme and hence in
order to keep things simple, in the following derivations we will drop
them completely).  The goal is to derive the 1PI four-point function 
$\ev{\imath\ov{q}_{\al}\imath q_{\ga}\imath\ov{q}_{\tau}\imath
q_{\eta}}$, and hence on the right hand side we will need, for
example, a term of the type
\be
\frac{\delta^3}{ \delta\imath q_{\eta t} \delta\imath \bar q_{\tau
w}\delta\imath q_{\ga z}} \ev{\imath\rho_{y}^{a}\imath\bar\chi_{\ba
x}}.
\nonumber
\ee
Functionally differentiating \eq{eq:rhoqq} with respect to $\imath
\bar q_{\tau w}$ and using the product rule, we obtain:
\bea
\frac{\delta^2}{\delta\imath \bar q_{\tau w}\delta\imath q_{\ga z}}
\ev{\imath\rho_{y}^{a}\imath\bar\chi_{\ba x}} &=&-S[\la,\ka]\left\{
\left[\frac{\delta}{\delta\imath \bar q_{\tau w}}
\ev{\imath\rho_{y}^{a}\imath J_{\la}}\right]
\ev{\imath\Phi_{\la}\imath q_{\ga z}\imath\Phi_{\ka}} \ev{\imath
J_{\ka} \imath\bar\chi_{\ba x}}\right.\nonumber\\
&&+
\ev{\imath\rho_{y}^{a}\imath J_{\la}} \ev{\imath\Phi_{\la}\imath \bar
q_{\tau w}\imath q_{\ga z}\imath\Phi_{\ka}} \ev{\imath J_{\ka}
\imath\bar\chi_{\ba x}}\nonumber\\
&&+
\eta_{\tau\ga}\eta_{\tau\ka}\left.  \ev{\imath\rho_{y}^{a}\imath
J_{\la}} \ev{\imath\Phi_{\la}\imath q_{\ga z}\imath\Phi_{\ka}}
\left[\frac{\delta}{\delta\imath \bar q_{\tau w}} \ev{\imath J_{\ka}
\imath\bar\chi_{\ba x}}\right]\right\}.
\eea
For clarity of presentation, we only retain the first term in the
product (and denote the rest with dots). Again, we make use of the
formula \eq{eq:mfd4} and obtain:
\be
\frac{\delta^2}{\delta\imath \bar q_{\tau w}\delta\imath q_{\ga z}}
\ev{\imath\rho_{y}^{a}\imath\bar\chi_{\ba x}}
=S[\la,\ka,\mu,\nu]
\ev{\imath\rho_{y}^{a}\imath J_{\mu}}
\ev{\imath\Phi_{\mu}\imath \bar q_{\tau w}\imath\Phi_{\nu}}
\ev{\imath J_{\nu}\imath J_{\la}}
\ev{\imath\Phi_{\la}\imath q_{\ga z}\imath\Phi_{\ka}}
\ev{\imath J_{\ka} \imath\bar\chi_{\ba x}}+\dots .
\ee
A last functional derivative with respect to the quark field 
$q_{\eta t}$ gives
\bea \frac{\delta^3}{\delta\imath q_{\eta t} \delta\imath \bar q_{\tau
w}\delta\imath q_{\ga z}} \ev{\imath\rho_{y}^{a}\imath\bar\chi_{\ba
x}} &=&S[\la,\ka,\mu,\nu]\nonumber\\
&&\times\left\{ \left[\frac{\delta}{\delta\imath q_{\eta t}}
\ev{\imath\rho_{y}^{a}\imath J_{\mu}} \right]
\ev{\imath\Phi_{\mu}\imath \bar q_{\tau w}\imath\Phi_{\nu}} \ev{\imath
J_{\nu}\imath J_{\la}} \ev{\imath\Phi_{\la}\imath q_{\ga
z}\imath\Phi_{\ka}} \ev{\imath J_{\ka} \imath\bar\chi_{\ba
x}}\right. \nonumber\\
&&+ \ev{\imath\rho_{y}^{a}\imath J_{\mu}} \ev{\imath\Phi_{\mu}\imath
q_{\eta t}\imath \bar q_{\tau w}\imath\Phi_{\nu}} \ev{\imath
J_{\nu}\imath J_{\la}} \ev{\imath\Phi_{\la}\imath q_{\ga
z}\imath\Phi_{\ka}} \ev{\imath J_{\ka} \imath\bar\chi_{\ba
x}}\nonumber\\
&&+\eta_{\eta\tau}\eta_{\tau\nu} \ev{\imath\rho_{y}^{a}\imath J_{\mu}}
\ev{\imath\Phi_{\mu}\imath \bar q_{\tau w}\imath\Phi_{\nu}}
\left[\frac{\delta}{\delta\imath q_{\eta t}} \ev{\imath J_{\nu}\imath
J_{\la}} \right] \ev{\imath\Phi_{\la}\imath q_{\ga z}\imath\Phi_{\ka}}
\ev{\imath J_{\ka} \imath\bar\chi_{\ba x}}\nonumber\\
&&+\eta_{\eta\tau} \ev{\imath\rho_{y}^{a}\imath J_{\mu}}
\ev{\imath\Phi_{\mu}\imath \bar q_{\tau w}\imath\Phi_{\nu}} \ev{\imath
J_{\nu}\imath J_{\la}} \ev{\imath\Phi_{\la}\imath q_{\eta t}\imath
q_{\ga z}\imath\Phi_{\ka}} \ev{\imath J_{\ka} \imath\bar\chi_{\ba
x}}\nonumber\\
&&+\eta_{\eta\tau}\eta_{\tau\ga}\eta_{\eta\ka} \left.
\ev{\imath\rho_{y}^{a}\imath J_{\mu}} \ev{\imath\Phi_{\mu}\imath \bar
q_{\tau w}\imath\Phi_{\nu}} \ev{\imath J_{\nu}\imath J_{\la}}
\ev{\imath\Phi_{\la}\imath q_{\ga z}\imath\Phi_{\ka}}
\left[\frac{\delta}{\delta\imath q_{\eta t}} \ev{\imath J_{\ka}
\imath\bar\chi_{\ba x}} \right]\right\}\nonumber\\
&&+\dots . 
\eea

\begin{figure}[t]
\vspace{0.5cm}
\includegraphics[width=1.0\linewidth]{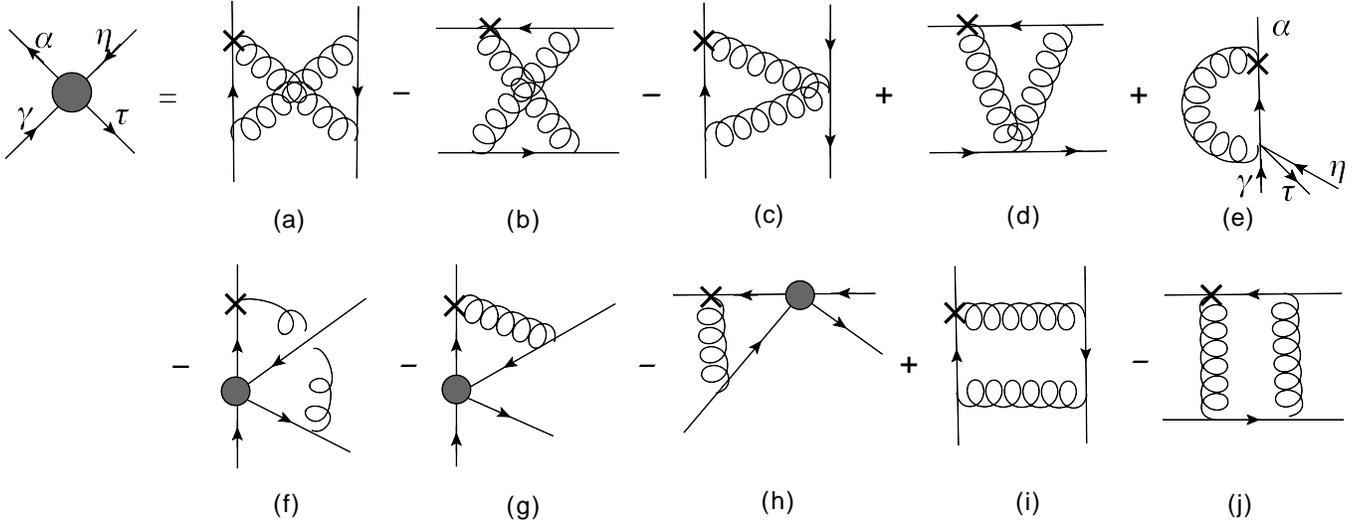}
\caption{\label{fig:1PI}
Diagrammatic representation of the \DS equation for the 1PI 4-point
quark-antiquark Green's function. Blobs represent dressed proper (1PI)
4-point vertex, solid lines represent the quark propagator, springs
denote either spatial ($\vec A$) or temporal ($\si$) gluon propagator
and cross denotes the tree level quark-gluon vertex. Internal
propagators and 1PI vertices are fully dressed.}
\end{figure}

Again, we take only the first term from the above sum and as before we
use the formula \eq{eq:mfd4}. We obtain:
\bea
\frac{\delta^3}{ \delta\imath  q_{\eta t}
\delta\imath \bar q_{\tau w}\delta\imath q_{\ga z}}
\ev{\imath\rho_{y}^{a}\imath\bar\chi_{\ba x}}
&=&-S[\la,\ka,\mu,\nu,\e,\de]
\ev{\imath\rho_{y}^{a}\imath J_{\e}}
\ev{\imath\Phi_{\e}\imath  q_{\eta t}\imath\Phi_{\de}}
\ev{\imath J_{\de}\imath J_{\mu}}\nonumber\\
&&\times
\ev{\imath\Phi_{\mu}\imath \bar q_{\tau w}\imath\Phi_{\nu}}
\ev{\imath J_{\nu}\imath J_{\la}}
\ev{\imath\Phi_{\la}\imath q_{\ga z}\imath\Phi_{\ka}}
\ev{\imath J_{\ka} \imath\bar\chi_{\ba x}}+\dots .
\eea
As for the gap equation, when sources/fields are set to zero, we can
identify the possible contributing field types denoted by internal
indices and determine $S[\la,\ka,\mu,\nu,\e,\de]$. Straightening out
the ordering of the external quark lines, the full \DS equation for
the proper (1PI) four-quark Green's function in configuration space
reads
\bea
\ev{\imath\ov{q}_{\al x}\imath q_{\ga z}\imath\ov{q}_{\tau w}
\imath q_{\eta t}}
&=&
[g\ga^0T^a]_{\al\ba}\int dy\, \de(x-y) \nonumber\\
&&\times\left\{ 
\left[ \ev{\imath\ov\chi_{\ba x}\imath\chi_{\ka}}
\ev{\imath\ov{q}_{\ka}\imath q_{\ga z}\imath \si^{c}_{\la}}
\ev{\imath\rho^c_{\la}\imath\rho^d_{\de}} \right] \left[
\ev{\imath\ov{q}_{\tau w}\imath q_{\nu}\imath \si^{b}_{\mu}}
\ev{\imath\ov\chi_{\nu}\imath\chi_{\e}} \ev{\imath\ov{q}_{\e}\imath
q_{\eta t}\imath \si^{d}_{\de}} \ev{\imath\rho^b_{\mu}\imath\rho^a_y}
\right]\right.\nonumber\\
&&-
\left[ \ev{\imath\ov\chi_{\ba x}\imath\chi_{\de}}
\ev{\imath\ov{q}_{\de}\imath q_{\eta t}\imath \si^{c}_{\e}}
\ev{\imath\rho^c_{\e}\imath\rho^d_{\ka}} \right] \left[
\ev{\imath\ov{q}_{\tau w}\imath q_{\nu}\imath \si^{b}_{\mu}}
\ev{\imath\ov\chi_{\nu}\imath\chi_{\la}} \ev{\imath\ov{q}_{\la}\imath
q_{\gamma z}\imath \si^{d}_{\ka}}
\ev{\imath\rho^b_{\mu}\imath\rho^a_y} \right]\nonumber\\
&&-
\left[ \ev{\imath\ov\chi_{\ba x}\imath\chi_{\ka}}
\ev{\imath\ov{q}_{\ka}\imath q_{\ga z}\imath \si^{c}_{\la}}
\ev{\imath\rho^c_{\la}\imath\rho^d_{\nu}} \ev{\ov{q}_{\tau w}\imath
q_{\eta t}\imath \si^{d}_{\nu}\imath \si^{b}_{\mu}}
\ev{\imath\rho^b_{\mu}\imath\rho^a_y} \right]\nonumber\\
&&+
\left[ \ev{\imath\ov\chi_{\ba x}\imath\chi_{\de}}
\ev{\imath\ov{q}_{\de}\imath q_{\eta t}\imath \si^{c}_{\e}}
\ev{\imath\rho^c_{\e}\imath\rho^d_{\ka}}\right]
\left[\ev{\imath\ov{q}_{\tau w}\imath q_{\ga z}\imath\si^d_{\ka}
\imath\si^b_{\la}} \ev{\imath\rho^b_{\la}\imath\rho^a_y}
\right]\nonumber\\
&&+
\left[ \ev{\imath\ov\chi_{\ba x}\imath\chi_{\ka}}
\ev{\imath\ov{q}_{\ka}\imath q_{\ga z}\imath\ov{q}_{\tau w}\imath
q_{\eta t}\si^{b}_{\la}} \ev{\imath\rho^b_{\la}\imath\rho^a_y}
\right]\nonumber\\
&&-
\left[ \ev{\imath\ov\chi_{\ba x}\imath\chi_{\ka}}
\ev{\imath\ov{q}_{\ka}\imath q_{\ga z}\imath\ov{q}_{\la}\imath q_{\eta
t}} \ev{\imath\ov{q}_{\tau w}\imath q_{\nu}\imath \si^{b}_{\mu}}
\ev{\imath\ov\chi_{\nu}\imath\chi_{\la}}
\ev{\imath\rho^b_{\mu}\imath\rho^a_y} \right]\nonumber\\
&&-
\left[ \ev{\imath\ov\chi_{\ba x}\imath\chi_{\ka}}
\ev{\imath\ov{q}_{\ka}\imath q_{\ga z}\imath\ov{q}_{\tau w}\imath
q_{\la}} \ev{\imath\ov\chi_{\la}\imath\chi_{\de}}
\ev{\imath\ov{q}_{\de}\imath q_{\eta t}\imath \si^{b}_{\e}}
\ev{\imath\rho^b_{\e}\imath\rho^a_y} \right]\nonumber\\
&&-
\left[ \ev{\imath\ov\chi_{\ba x}\imath\chi_{\nu}}
\ev{\imath\ov{q}_{\nu}\imath q_{\mu}\imath\ov{q}_{\tau w}\imath
q_{\eta t}} \ev{\imath\ov\chi_{\mu}\imath\chi_{\ka}}
\ev{\imath\ov{q}_{\ka}\imath q_{\ga z}\imath \si^{b}_{\la}}
\ev{\imath\rho^b_{\la}\imath\rho^a_y} \right]\nonumber\\
&&+\left[
\ev{\imath\ov\chi_{\ba x}\imath\chi_{\ka}}
\ev{\imath\ov{q}_{\ka}\imath q_{\ga z}\imath \si^{c}_{\la}}
\ev{\imath\rho^c_{\la}\imath\rho^d_{\nu}} \right] \left[
\ev{\imath\ov{q}_{\tau w}\imath q_{\mu}\imath \si^{d}_{\nu}}
\ev{\imath\ov\chi_{\mu}\imath\chi_{\de}} \ev{\imath\ov{q}_{\de}\imath
q_{\eta t}\imath \si^{b}_{\e}} \ev{\imath\rho^b_{\e}\imath\rho^a_y}
\right]\nonumber\\
&&-
\left.  \left[ \ev{\imath\ov\chi_{\ba x}\imath\chi_{\de}}
\ev{\imath\ov{q}_{\de}\imath q_{\eta t}\imath \si^{c}_{\e}}
\ev{\imath\rho^c_{\e}\imath\rho^d_{\nu}} \right] \left[
\ev{\imath\ov{q}_{\tau w}\imath q_{\mu}\imath \si^{d}_{\nu}}
\ev{\imath\ov\chi_{\mu}\imath\chi_{\ka}} \ev{\imath\ov{q}_{\ka}\imath
q_{\ga z}\imath \si^{b}_{\la}} \ev{\imath\rho^b_{\la}\imath\rho^a_y}
\right] \right\} \nonumber\\
&&+\dots\label{eq:1pi_4quark}
\eea
where the dots represent the $\vec A$ vertex terms which are not
considered here and we have already replaced the tree-level temporal
quark-gluon vertex with its expression \eq{eq:treelevelquarkvertex1}.
The expression \eq{eq:1pi_4quark} is diagrammatically represented in
\fig{fig:1PI} and the terms have been reordered such that the first
term corresponds to the diagram (a) of \fig{fig:1PI}, the second term
corresponds to diagram (b) and so on. The term presented explicitly
above corresponds to diagram (i).  Also, notice the minus sign in the
term corresponding to diagram (h) --- this is the origin of the minus
sign arising in the kernel of the homogeneous \BS equation considered
in Ref.~\cite{Popovici:2010mb}.

The next step is to derive the connection between the amputated
connected 4-point quark-antiquark Green's function (used
conventionally in the \BS equation), and the 1PI Green's function.
Starting with the identity
\be
\ev{\imath q_{\ba}\imath\ov{q}_{\ga}}\ev{\imath\chi_{\ga}
\imath\ov\chi_{\al}}=\de_{\ba\al},
\ee
we take two subsequent functional derivatives with respect to generic
sources, $\imath J_{\de}$ and $\imath J_{\la}$. Using \eq{eq:mfd11}
recursively, choosing $\de,\la$ as needed, we eventually obtain the
standard expression (in terms of our notation and conventions) for the
4-point quark-antiquark connected Green's function, written in terms
of 1PI Green's functions (and omitting the $\vec A$-vertex
contributions, which will not play a role here):
\bea
\ev{\imath\ov\chi_{\al}\imath\chi_{\de}\imath\ov
\chi_{\la}\imath\chi_{\eta}}&=&
\ev{\imath\ov\chi_{\al}\imath\chi_{\ga}}
\ev{\imath\ov\chi_{\la}\imath\chi_{\e}} \ev{\imath\ov{q}_{\ga}
\imath q_{\ka}\imath\ov{q}_{\e}\imath q_{\ba}}
\ev{\imath\ov\chi_{\ka}\imath\chi_{\de}}
\ev{\imath\ov\chi_{\ba}\imath\chi_{\eta}}
\nonumber\\
&&+
\left[
\ev{\imath\ov\chi_{\al}\imath\chi_{\ga}}
\ev{\imath\ov{q}_{\ga}\imath q_{\ba}\imath \si_{\ka}^a}
\ev{\imath\ov\chi_{\ba}\imath\chi_{\eta}}
\right]
\left[
\ev{\imath\ov\chi_{\la}\imath\chi_{\tau}}
\ev{\imath\ov{q}_{\tau}\imath q_{\mu}\imath \si_{\nu}^b}
\ev{\imath\ov\chi_{\mu}\imath\chi_{\de}}
\right]
\ev{\imath \rho_{\ka}^a\imath \rho_{\nu}^b}\nonumber\\
&&-
\left[
\ev{\imath\ov\chi_{\al}\imath\chi_{\tau}}
\ev{\imath\ov{q}_{\tau}\imath q_{\mu}\imath \si_{\nu}^b}
\ev{\imath\ov\chi_{\mu}\imath\chi_{\de}}
\right]
\left[
\ev{\imath\ov\chi_{\la}\imath\chi_{\e}}
\ev{\imath\ov{q}_{\e}\imath q_{\ba}\imath \si_{\ka}^a}
\ev{\imath\ov\chi_{\ba}\imath\chi_{\eta}}
\right]
\ev{\imath \rho_{\ka}^a\imath \rho_{\nu}^b}.
\label{eq:connected_4quark}
\eea

\begin{figure}[t]
\vspace{0.5cm}
\includegraphics[width=0.6\linewidth]{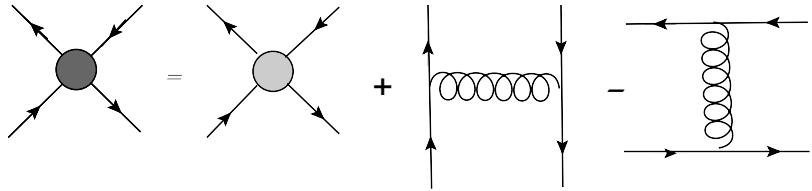}
\caption{\label{fig:1PI_amputated}
Relation between the 1PI (dark blob) and amputated (shaded blob)
4-point Greens function for the quark-antiquark system. Internal
propagators and 1PI vertices are fully dressed.}
\end{figure}

The relation \eq{eq:connected_4quark} can be further simplified by
introducing the fully amputated Green's function, i.e.  dividing by
the quark propagators (cut the quark legs), as shown in
\fig{fig:1PI_amputated}.  Before we consider the explicit form of Eqs.
(\ref{eq:1pi_4quark}, \ref{eq:connected_4quark}), it is necessary to
briefly recall the heavy mass expansion and the truncation scheme that
will be applied in this study. This is the topic of the next section.

\section{Heavy quark mass expansion}
\label{sec:hqme}

Consider again the quark contribution to the full QCD generating
functional, \eq{eq:genfunc}.  In the following, we briefly describe
the derivation of the quark (and antiquark) propagator, in the heavy
mass limit and under the truncation of the \YM sector to include only
the dressed two-point functions.  The presentation follows the
original work of Ref.~\cite{Popovici:2010mb}.

The full quark field is decomposed according to the heavy quark
transformation
\bea
q_\al(x)=e^{-\imath mx_0}\left[h(x)+H(x)\right]_\al,\,
&h_\al(x)=e^{\imath mx_0}\left[P_+q(x)\right]_\al,\,
&H_\al(x)=e^{\imath mx_0}
\left[P_-q(x)\right]_\al
\label{eq:qdecomp}
\eea
(similarly for the antiquark field), where the two components $h$ and
$H$ are introduced with the help of the spinor projectors
\be
P_\pm=\frac{1}{2}(\openone\pm\ga^0).
\ee
This corresponds to a particular case of the heavy quark transform
underlying HQET \cite{Neubert:1993mb}, but in the functional approach
considered here it is simply an arbitrary decomposition that will turn
out to be very useful in Coulomb gauge. As will shortly be discussed,
this will lead to the suppression of the spatial gluon propagator and
spatial quark-gluon vertex at leading order in the mass expansion.

The quark fields, decomposed according to \eq{eq:qdecomp}, are now
inserted into the generating functional \eq{eq:genfunc}. Further, the
$H$-fields are integrated out, and an expansion in the heavy quark
mass is performed (throughout this work, we will follow the usual
terminology and use the expression ``mass expansion'', instead of
``expansion in the \emph{inverse} mass''). At leading order, the
generating functional reads:
\bea
Z[\ov{\chi},\chi]&=&\int{\cal D}\Phi\exp{\left\{\imath
\int d^4x\ov{h}_\al(x)
\left[\imath\partial_{0x}+gT^a\si^a(x)\right]_{\al\ba}
h_\ba(x)\right\}}
\nonumber\\&&\times
\exp{\left\{\imath
\int d^4x\left[e^{-\imath mx_0}\ov{\chi}_\al(x) h_\al(x)
+e^{\imath mx_0}\ov{h}_\al(x)\chi_\al(x)\right]+
\imath{\cal S}_{YM}\right\}}+{\cal O}\left(1/m\right),
\label{eq:genfunc4}
\eea
where the temporal component of the covariant derivative $D_0$ has
been written explicitly. Notice that, as a consequence of the
decomposition \eq{eq:qdecomp}, the spin degrees of freedom have
decoupled from the system.  Importantly, in the above we have kept the
full quark and antiquark sources (rather than the ones corresponding
to the large components of the quark field $h$, as is usually done in
HQET). This means that the methods of functional formalism are
preserved at leading order in the mass expansion, and in turn this
implies that we are allowed to use the full \DS equations of QCD,
i.e. the gap and 4-point \DS equations of the last section, replacing,
however, the kernels, propagators and vertices by their leading order
in the mass expansion.

The gap equation \eq{eq:gap} is solved together with the \ST identity.
This is derived from the invariance of the QCD action under a
time-dependent Gauss-BRST transform \cite{Popovici:2010mb}, and in
Coulomb gauge it has the following expression:
\bea
k_3^0\G_{\ov{q}q\si\al\ba}^{d}(k_1,k_2,k_3)&=&
\imath\frac{k_{3i}}
{\vec{k}_3^2}\G_{\ov{q}qA\al\ba i}^{a}(k_1,k_2,k_3)
\G_{\ov{c}c}^{ad}(-k_3)\nonumber\\&&
+\G_{\ov{q}q\al\de}(k_1)\left[\tilde{\G}_{\ov{q};\ov{c}cq}^{d}
(k_1+q_0,k_3-q_0;k_2)+\imath gT^d\right]_{\de\ba}
\nonumber\\&&
+\left[\tilde{\G}_{q;\ov{c}c\ov{q}}^{d}(k_2+q_0,k_3-q_0;k_1)-
\imath gT^d\right]_{\al\de}\G_{\ov{q}q\de\ba}(-k_2)
\label{eq:stid}
\eea
where $k_1+k_2+k_3=0$, $q_0$ is an arbitrary energy injection scale
(arising from the noncovariance of Coulomb gauge
\cite{Watson:2008fb}), $\G_{\ov{c}c}$ is the ghost proper two-point
function, $\tilde{\G}_{\ov{q};\ov{c}cq}$ and
$\tilde{\G}_{q;\ov{c}c\ov{q}}$ are ghost-quark kernels associated with
the Gauss-BRST transform.

Since in the generating functional \eq{eq:genfunc4} the tree-level
spatial quark gluon vertex $\G_{\ov{q}qA}^{(0)}$ appears at $O(1/m)$
in the mass expansion (which is here neglected), and furthermore,
under our truncation the pure \YM vertices are also neglected, it
follows that the \DS equation for the nonperturbative spatial
quark-gluon vertex provides the result that $\G_{\bar qqA}\sim O(1/m)$
(see \cite{Popovici:2010mb} for a complete discussion and
justification of this truncation).  Moreover, the ghost-quark kernels
involve pure \YM vertices and can be also neglected. Thus, under our
truncation scheme, the \ST identity reduces to
\be
k_3^0\G_{\ov{q}q\si\al\ba}^{d}(k_1,k_2,k_3)=
\G_{\ov{q}q\al\de}(k_1)\left[\imath gT^d\right]_{\de\ba}-
\left[\imath gT^d\right]_{\al\de}\G_{\ov{q}q\de\ba}(-k_2)
+{\cal O}\left(1/m\right).
\label{eq:stidlimit}
\ee
We then insert this into \eq{eq:gap}, together with the tree-level
quark proper two-point function
\be
\G_{\ov{q}q\al \ba}^{(0)}(k)=\imath\de_{\al\ba}\left[k_0-m\right]
+{\cal O}\left(1/m\right)
\label{eq:gaptree}
\ee
and the tree level quark gluon vertex 
\be
\G_{\ov{q}q\si\al\ba}^{(0)a}(k_1,k_2,k_3)=
\left[gT^a\right]_{\al\ba}+{\cal O}\left(1/m\right)
\label{eq:feyn0}
\ee
that follow from the generating functional \eq{eq:genfunc4}. The
nonperturbative temporal gluon propagator reads \cite{Watson:2007vc}:
\be
W_{\si\si}^{ab}(k)=\de^{ab} W_{\si\si}(\vec k)
=\de^{ab}\frac{\imath}{\vec{k}^2}D_{\si\si}(\vec{k}^2).
\label{eq:Wsisi}
\ee
Lattice results \cite{Quandt:2008zj} motivate that the dressing
function $D_{\si\si}$ is largely independent of energy and moreover,
$D_{\si\si}$ is infrared divergent and likely to behave as
$1/\vec{k}^2$ for vanishing $\vec{k}^2$ (we will only use the explicit
form of $D_{\si\si}$ in the last step of the calculation). Putting all
this together, we find the following solution to \eq{eq:gap} for the
heavy quark propagator:
\bea
W_{\ov{q}q\al\ba}(k)=\frac{-\imath\de_{\al\ba}}{\left[k_0-m-
{\cal I}_r+\imath\e\right]}+{\cal O}\left(1/m\right),
\label{eq:quarkpropnonpert}
\eea
with the constant
[$\dk{\vec{\w}}=d^3\vec{\w}/(2\pi)^3$]
\be
 {\cal I}_r =\frac{1}{2}g^2C_F
\int_r\frac{\dk{\vec{\w}}D_{\si\si}(\vec{\w})}{\vec{\w}^2}
+{\cal O}\left(1/m\right).
\label{eq:iregularized}
\ee
 The (implicit) regularization of $ {\cal I}_r$ is signaled by the
symbol ``$r$''. When solving \eq{eq:gap}, the ordering of the
integration is set such that the temporal integral is performed first,
under the condition that the spatial integral is regularized and
finite.  With the solution \eq{eq:quarkpropnonpert}, we return to the
\ST identity, \eq{eq:stidlimit}, and find that nonperturbatively the
temporal quark-gluon vertex remains bare:
\be
\G_{\ov{q}q\si\al\ba}^{a}(k_1,k_2,k_3)=
\left[gT^a\right]_{\al\ba}+{\cal O}\left(1/m\right).
\label{eq:feyn}
\ee

Let us now briefly review the properties of the propagator
\eq{eq:quarkpropnonpert} (see also \cite{Popovici:2010mb}).  As
opposed to the conventional QCD quark propagator, which possesses a
pair of simple poles, in this case we only have a single pole in the
complex $k_0$-plane, and the explicit Feynman prescription becomes
important.  It then follows that the closed quark loops (constructed
from virtual quark-antiquark pairs) vanish due to the energy
integration, and this implies that the theory is quenched in the heavy
mass limit:
\be
\int\frac{dk_0}{\left[k_0-m- {\cal I}_r+\imath\e\right]
\left[k_0+p_0-m- {\cal I}_r+\imath\e\right]}=0.
\label{eq:tempint}
\ee
Notice also that the propagator \eq{eq:quarkpropnonpert} is diagonal
in the outer product of the fundamental color, flavor and spinor
spaces.  This corresponds to the decoupling of the spin from the heavy
quark system, due to the decomposition \eq{eq:qdecomp}.  Indeed,
$W_{\ov{q}q}^{(0)}$ differs from the heavy quark tree-level propagator
\cite{Neubert:1993mb} only through the mass term, and this is due to
the fact that we retain the sources of the full quark fields, while in
HQET one uses the sources for the large $h$-fields directly.  Finally,
we also stress that the position of the pole has no physical meaning
(the quark can never be on-shell). Since the poles in the quark
propagator are situated at infinity (this becomes apparent after the
regularization in ${\cal I}_r$ is removed assuming, as indicated by
the lattice \cite{Quandt:2008zj}, that the temporal gluon propagator
is infrared enhanced), one requires infinite energy to create a single
quark from the vacuum or, if a hadronic system is considered, only the
relative energy (derived from the \BS equation) is important. Indeed,
it is precisely the exact cancellation of the divergent constants that
lead to the separation of physical and unphysical poles.

As discussed in Ref.~\cite{Popovici:2010mb} (and visible in the single
pole structure of the quark propagator), the mass expansion leads to
the separation of the quark and antiquark states. Whilst virtual
$q\bar q$-loops cancel due to the energy integration, \eq{eq:tempint},
physical bound states for $\bar qq$ pairs should still exist. This
means that we have to consider the gap equation for the antiquark
propagator (and its Feynman prescription) separately.  The gap
equation for the antiquark propagator is derived from the full QCD gap
equation \eq{eq:gap}, by reversing the ordering of the quark and
antiquark functional derivatives that lead to the quark Green's
functions:
\bea
-\G_{q\ov{q}\de\al}(-k)&=-\G_{q\ov{q}\de\al}^{(0)}(-k)-\int\dk{\w}&
\left\{\G_{q\ov{q}\si\de\ga}^{b}(-k,\w,k-\w)W_{q\ov{q}\ga\ba}(-\w)
\G_{q\ov{q}\si\ba\al}^{(0)a}(-\w,k,\w-k)W_{\si\si}^{ab}(k-\w) \right.
\nonumber\\&&\left.
+\G_{q\ov{q}A\de\ga j}^{b}(-k,\w,k-\w)W_{q\ov{q}\ga\ba}(-\w)\G_{q
\ov{q}A\ba\al i}^{(0)a}(-\w,k,\w-k)W_{AAij}^{ab}(k-\w)
\right\}.\nonumber\\
\label{eq:bgap}
\eea
Again, we apply our truncation scheme and the above equation reduces
to:
\be
\G_{q\ov{q}\de\al}(-k)=\G_{q\ov{q}\de\al}^{(0)}(-k)+\int\dk{\w}\G_{
q\ov{q}\si\de\ga}^{b}(-k,\w,k-\w)W_{q\ov{q}\ga\ba}(-\w)\G_{q\ov{q}
\si\ba\al}^{(0)a}(-\w,k,\w-k)W_{\si\si}^{ab}(k-\w)+{\cal O}
\left(1/m\right).
\label{eq:bglap1}
\ee
In similar manner, we derive the \ST identity for the antiquark-gluon
vertex:
\be
-k_3^0\G_{q\ov{q}\si\ba\al}^{d}(k_2,k_1,k_3)=
+\G_{q\ov{q}\ba\de}(k_2)\left[\imath
gT^d\right]_{\de\al}^T-\left[\imath gT^d
\right]_{\ba\de}^T\G_{q\ov{q}\de\al}(-k_1)+{\cal O}\left(1/m\right).
\ee
The following antiquark propagator is obtained, as a solution to
\eq{eq:bglap1}:
\bea
W_{q\ov{q} \al\ba}(k)=\frac{-\imath \de_{\al\ba}} {\left[k_0+m-{\cal
I}_r+\imath\e\right]} +{\cal O}\left(1/m\right)
\label{eq:antiquarkpropnonpert}.
\eea
The corresponding temporal antiquark-gluon vertex is given by:
\be
\G_{q\ov{q}\si \al\ba}^{a}(k_1,k_2,k_3)=
-\left[gT^a\right]_{\ba\al}+{\cal O}\left(1/m\right).
\label{eq:antiqqsinp}
\ee
In the above, notice the Feynman prescription of the propagator, as
well as the sign of the loop correction.  This will have the
consequence that the \BS equation for the quark-antiquark states can
be interpreted as a bound state equation. There, the quark and the
antiquark do not create a virtual quark-antiquark pair (as in the case
of the closed quark loops), but a system composed of two separate
unphysical particles (i.e., they are not connected by a single
primitive vertex). Moreover, in the \BS equation the so-called crossed
box contributions (i.e., nonplanar diagrams that contain any
combinations of nontrivial interactions allowed within our truncation
scheme) cancel -- in fact, this type of cancellation will also appear
in the 4-point Green's function. This is due to the temporal
integration performed over multiple propagators with the same relative
sign for the Feynman prescription (similar to \eq{eq:tempint}, but in
this case the terms originate from internal quark or antiquark
propagators). As a consequence, the \BS kernel reduces to the ladder
truncation \cite{Popovici:2010mb}.

\section{Solution in the heavy mass limit}

Having reviewed the properties of the various vertices and propagators
under truncation and in the heavy mass limit, we can now return to the
formula for the 4-point Green's function \eq{eq:1pi_4quark} (combined
with \eq{eq:connected_4quark}), and the corresponding diagrammatic
representations \fig{fig:1PI}, \fig{fig:1PI_amputated}, and apply our
truncation scheme in the heavy quark limit, at leading order in the
mass expansion.

\subsection{One particle irreducible Green's function}

For the quark-antiquark system, we consider the flavor non-singlet
Green's function in the $s$-channel, where the quark and the antiquark
are regarded as two distinct flavors (but with equal masses).  Hence,
the diagrams (a), (c) and (i) of \fig{fig:1PI} are excluded.
\begin{figure}[t]
\vspace{0.5cm}
\centering\includegraphics[width=0.27\linewidth]{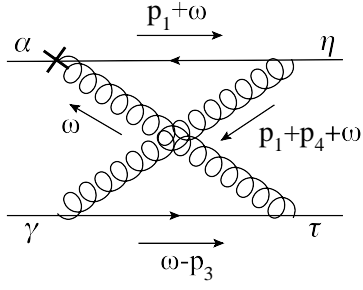}
\caption{\label{fig:diagr-b}
Crossed ladder diagram that contributes to the 1PI 4-point Green's
function. The upper line denotes the quark propagator, the lower one,
the antiquark propagator and springs denote the temporal gluon
propagator.}
\end{figure}
The diagram (b) (crossed ladder type exchange diagram) explicitly
reads in momentum space and without the minus sign prefactor (see also
\fig{fig:diagr-b}):
\bea
\lefteqn{\int\dk{\w}\left[\G_{\bar qq\si
\al\de}^{(0)a}(p_1,-p_1-\w,\w)W_{\bar qq \de\phi}(p_1+\w) \G_{\bar
qq\si \phi\eta}^{b}(p_1+\w,p_4,-p_1-p_4-\w)\right]}\nonumber\\
&&\times\left[\G_{\bar qq\si\tau\mu}^{c}(p_3,\w-p_3,-\w) W_{\bar
qq\mu\la}(p_3-\w) \G_{\bar
qq\si\la\ga}^{d}(p_3-\w,p_2,p_1+p_4+\w)\right] W_{\si\si}^{ac}(-\w)
W_{\si\si}^{bd}( p_1+ p_4+\w)
\label{eq:diagr-b}
\eea
The above expression refers to the full QCD case (but omitting the
spatial $W_{AA}$ contributions). The lower line corresponds to the
antiquark, so we first rewrite it in terms of the (full QCD) antiquark
propagators and vertices, using $W_{\bar qq
\mu\la}(p_3-\w)=\left[-W_{q\bar q}^{T}(\w-p_3)\right]_{\mu\la}$ and
$\G_{\bar qq\si \tau\mu}=\left[-\G_{q\bar
q\si}^T\right]_{\tau\mu}$. Then we apply the mass expansion by
inserting the appropriate propagators Eqs. (\ref{eq:quarkpropnonpert},
\ref{eq:antiquarkpropnonpert}) and vertices Eqs. (\ref{eq:feyn},
\ref{eq:antiqqsinp}). Noticing the energy independence of the temporal
gluon propagator, one sees then immediately that the energy integral
vanishes, due to the fact that both quark and antiquark propagators
have the same Feynman prescription relative to $\w_0$:
\be
\int\dk{\w_{0}}W_{\bar qq }(p_1+\w) W_{q\bar q}(\w-p_3)\sim
\int\dk{\w_{0}} \left\{ \frac{1 }{\w_0+p_1^0-m-{\cal I}_r+\imath\e}
-\frac{1}{\w_0-p_3^0+m-{\cal I}_r+\imath\e} \right\}=0.
\label{eq:diagr-b-energy}
\ee
That this crossed ladder diagram vanishes within the scheme considered
here is exactly the analogous to the absence of the crossed ladder
contribution to the kernel of the homogeneous \BS equation in
Ref.~\cite{Popovici:2010mb}.

Turning to diagram (d), this contribution involves the $\G_{\bar
 qq\si\si}$ vertex. In the Appendix it is shown that under the scheme
considered here, the \ST identity furnishes the result that the
nonperturbative quark-2 gluon vertex (and hence the diagram (d))
vanishes. Again, this result is analogous to the discussion of the
kernel in the homogeneous \BS equation from \cite{Popovici:2010mb}.

Let us for the moment discard the diagrams (f) and (g), which include
the 1PI 4-point quark Green's function, and the diagram (e),
containing a 4 quark-gluon vertex.  Then we are only left with the
diagram (h) and the rainbow-ladder term (j). This simplification
enables us to derive a solution for the corresponding (truncated)
equation for the 1PI 4-point quark Green's function. With this result
at hand, we will then return to the diagrams (f), (g) and (e), and
explicitly show that they cancel (and hence our assumption is
justified).  With these observations, \eq{eq:1pi_4quark} reduces to:
\bea
\ev{\imath\ov{q}_{\al x}\imath q_{\ga z}\imath\ov{q}_{\tau w}
\imath q_{\eta t}}
&=&- \left[g T^{a}\ga^{0}\right] _{\al\ba}
\int dy \de(x-y)
\left\{
\left[
\ev{\imath\ov\chi_{\ba x}\imath\chi_{\de}}
\ev{\imath\ov{q}_{\de}\imath q_{\eta t}\imath \si^{c}_{\e}}
\ev{\imath\rho^c_{\e}\imath\rho^d_{\nu}}
\right]\right.\nonumber\\
&&\times\left[
\ev{\imath\ov{q}_{\tau w}\imath q_{\mu}\imath \si^{d}_{\nu}}
\ev{\imath\ov\chi_{\mu}\imath\chi_{\ka}}
\ev{\imath\ov{q}_{\ka}\imath q_{\ga z}\imath \si^{b}_{\la}}
\ev{\imath\rho^b_{\la}\imath\rho_y^a}
\right]
\nonumber\\
&&+
\left.
\ev{\imath\ov\chi_{\ba x}\imath\chi_{\nu}}
\ev{\imath\ov{q}_{\nu}\imath q_{\mu}\imath\ov{q}_{\tau w}
\imath q_{\eta t}}
\ev{\imath\ov\chi_{\mu}\imath\chi_{\ka}}
\ev{\imath\ov{q}_{\ka}\imath q_{\ga z}\imath \si^{b}_{\la}}
\ev{\imath\rho^b_{\la}\imath\rho_{y}^a}
\right\}. 
\label{eq:1pi_4quark1}
\eea

\begin{figure}[t]
\vspace{0.5cm}
\centering\includegraphics[width=0.8\linewidth]{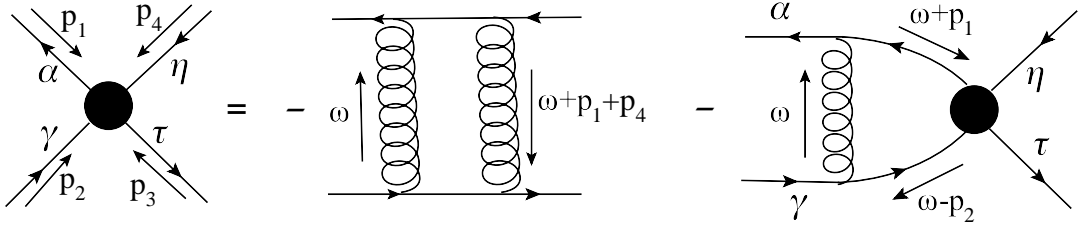}
\caption{\label{fig:1PI_mr}
Truncated \DS equation for the 1PI 4-point Green's function in the
$s$-channel. Same conventions as in \fig{fig:1PI} apply.}
\end{figure}
It is convenient to express the resulting equation in momentum space.
We define
\be
\ev{\imath\ov{q}_{\al x}\imath q_{\ga z}\imath\ov{q}_{\tau w}\imath
q_{\eta t}}= \int \dk{k_1}\dk{k_2}\dk{k_3}\dk{k_4} e^{-\imath k_1\cdot
x-\imath k_2\cdot z-\imath k_3\cdot w-\imath k_4\cdot t}
\G^{(4)}_{\al\ga\tau\eta}(k_1,k_2,k_3,k_4)(2\pi)^{4}
\de(k_1+k_2+k_3+k_4)
\label{eq:g4fourier}
\ee
and arrive at the following \DS equation for the 1PI 4-point quark
Green's function in the $s$-channel (shown diagrammatically in
\fig{fig:1PI_mr}):
\bea
\lefteqn{ \G^{(4)}_{\al\ga\tau\eta}(p_1,p_2,p_3,p_4)=
-\int\dk{\w}\left[\G_{\bar qq\si \al\ba}^{(0)a}(p_1,-p_1-\w,\w)W_{\bar
qq \ba\de}(p_1+\w) \G_{\bar qq\si
\de\eta}^{c}(p_1+\w,p_4,-p_1-p_4-\w)\right]}
\nonumber\\
&&\times\left[\G_{\bar qq\si\tau\mu}^{d}(p_3,p_2-\w,p_1+p_4+\w)
W_{\bar qq\mu\ka}(\w-p_2) \G_{\bar
qq\si\ka\ga}^{b}(\w-p_2,p_2,-\w)\right] W_{\si\si}^{ab}(- \w)
W_{\si\si}^{cd}(p_1+ p_4+\w)
\nonumber\\
&&-\int\dk{\w}\G_{\bar qq\si\al\ba}^{(0)a} (p_1,-p_1-\w,\w) W_{\bar qq
\ba\nu}(p_1+\w) \G^{(4)}_{\nu\mu\tau\eta}(p_1+\w,p_2-\w,p_3,p_4)
W_{\bar qq\mu\ka}(\w-p_2)\G_{\bar qq\si \ka\ga}^{b} (\w-p_2,p_2,-\w)
\nonumber\\
&&\times W_{\si\si}^{ab}(-\w).
\label{eq:4pointDS1}
\eea

In order to proceed, we make the following assumption for the energy
and momentum dependence of the function $\G^{(4)}$:
\be
\G ^{(4)}(p_1,p_2,p_3,p_4)=\G ^{(4)}(P_0;\vec p_1+\vec p_4),
\label{eq:ansatzG4}
\ee
with $P_0=p_1^0+p_2^0$. This implies that in the above equation the
4-point function $\G^{(4)}_{\nu\mu\tau\eta}(p_1+\w,p_2-\w,p_3,p_4) $
does not depend on the integration variable $\w_0$, and hence we can
separate the energy and three-momentum integrals. Moreover, with this
Ansatz we will be allowed to Fourier transform the resulting spatial
integral back to coordinate space, as shall be explained shortly
below.

We identify the antiquark component of \eq{eq:4pointDS1} (lower line
of \fig{fig:1PI_mr}) and insert the expressions
Eqs. (\ref{eq:quarkpropnonpert}, \ref{eq:antiquarkpropnonpert}), for
the quark and antiquark propagators, along with the vertices
Eqs. (\ref{eq:feyn}, \ref{eq:antiqqsinp}).  We also note the Fierz
identity for the generators
\be
2\left[T^a\right]_{\al\ba}\left[T^a\right]_{\de\ga}=\de_{\al\ga}
\de_{\de\ba}-\frac{1}{N_c}\de_{\al\ba}\de_{\de\ga}.
\label{eq:fierz}
\ee
 After completing the energy integration and with the definition
\eq{eq:Wsisi} for the temporal gluon propagator, \eq{eq:4pointDS1}
simplifies to :
\bea
\left[ P_0-2 {\cal I}_{r} +2\imath\e\right]
\G^{(4)}_{\al\ga\tau\eta}(P_0;\vec p_1+\vec p_4) 
&=&\imath\frac{g^4}{4}\left[\left(N_c-\frac{2}{N_c}\right)
\de_{\al\ga}\de_{\tau\eta}
+\frac{1}{N_c^2}\de_{\al\eta}\de_{\tau \ga}\right]
\int\dk{\vec\w}W_{\si\si}(\vec\w)W_{\si\si}(\vec p_1+\vec
p_4+\vec\w)\nonumber\\
&&+\imath
\frac{g^2}{2}\left[\de_{\al\ga}\de_{\mu\nu}
-\frac{1}{N_c}\de_{\al\nu}\de_{\mu\ga}\right]
\int\dk{\vec\w}W_{\si\si}(\vec\w) \G^{(4)}_{\nu\mu\tau\eta}
(P_0;\vec p_1+\vec p_4+\vec \w). \nonumber\\
\label{eq:4pointDSen}
\eea

Let us now make the following color decomposition for the function
$\G^{(4)}$:
\be
\G^{(4)}_{\al\ga\tau\eta}
=\de_{\al\ga}\de_{\tau\eta}\G_{1}^{(4)}+\de_{\al\eta}\de_{\tau
\ga}\G_{2}^{(4)}.
\label{eq:g4color1PI}
\ee
where $\G^{(4)}_{1}$ and $\G^{(4)}_{2}$ are scalar functions.
\begin{figure}[t]
\vspace{0.5cm}
\centering\includegraphics[width=0.25\linewidth]{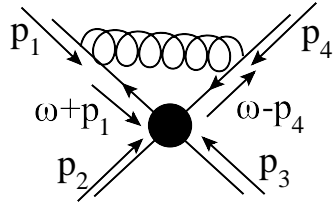}
\caption{\label{fig:diagr-g}
Momentum routing for the diagram (g). See text for details.}
\end{figure}
At this point, it is convenient to Fourier transform back to
coordinate space.  In general, since \eq{eq:4pointDS1} might in
principle contain momentum-dependent vertex functions, as well as
mixing of energy and three-momentum variables, this transformation
could not be carried out. However, in our case (\eq{eq:4pointDSen})
momentum-dependent vertices are absent and moreover, with the Ansatz
\eq{eq:ansatzG4}, the energy and three-momentum integrals have
separated such that the spatial integral is performed only over two
functions (spatial gluon propagator and the spatial component of the
quark 4-point function, both functions of momentum squared). Hence the
Fourier transform simplifies to:
\bea
\int\dk{\vec \w} W_{\si\si}(\vec \w^2)\G^{(4)}(P_0;(\vec q+\vec
\w)^2)= \int d \vec x e^{-\imath\vec q\cdot\vec x}
W_{\si\si}(x)\G^{(4)}(P_0; x),
\label{eq:convolution}
\eea
with $x=|\vec x|$. Sorting out the color factors, it is
straightforward to obtain for the components $\G_1^{(4)}$, $\G_2^
{(4)}$:
\bea
\G_1^{(4)}(P_0; x)&=& \imath \left(\frac{g^2}{2N_c}\right)^2
\frac{W_{\si\si}(x)^2 N_c \left[(P_0-2{\cal I}_r)(N_c^2-2)+\imath g^2
C_F W_{\si\si}(x)\right]} {[P_0-2{\cal I}_r+\imath
\frac{g^2}{2N_c}W_{\si\si}(x)+2\imath\e] [P_0-2{\cal I}_r-\imath
g^2C_FW_{\si\si}(x) +2\imath\e]},
\nonumber\\
\G_2^{(4)}(P_0; x)&=&\imath\left(\frac{g^2}{2N_c}\right)^2
\frac{W_{\si\si}(x)^2} {P_0-2{\cal I}_r+\imath
\frac{g^2}{2N_c}W_{\si\si}(x) +2\imath\e},
\eea
where $x$ is the separation associated with the momentum $\vec
p_1+\vec p_4$.  Inserting the above results into the decomposition
\eq{eq:g4color1PI}, we find the final formula for the 1PI quark
Green's function:
\bea
\G^{(4)}_{\al\ga\tau\eta} (P_0; x)
&=&\imath\left(\frac{g^2}{2N_c}\right)^2 \frac{W_{\si\si}(x)^2}
{P_0-2{\cal I}_r+\imath \frac{g^2}{2N_c}W_{\si\si}(x)
+2\imath\e}\nonumber\\
&&\times\left\{ \de_{\al\ga}\de_{\tau\eta} \frac{(P_0-2{\cal
I}_r)N_c(N_c^2-2)+\imath g^2N_c C_F W_{\si\si}(x)} {P_0-2{\cal
I}_r-\imath g^2C_FW_{\si\si}(x) +2\imath\e}
+\de_{\al\eta}\de_{\tau\ga} \right\}.
\label{eq:sol1PI}
\eea

Having derived the solution \eq{eq:sol1PI} for the 1PI Green's
function, we return to the diagrams (f), (g) and (e) and show that
they do not contribute to the final result. To see this, we first
consider the diagram (g) and study the energy integral (see also
\fig{fig:diagr-g}).  Noticing that the energy dependence of the
internal four-point function can be written as
\be
\G^{(4)}(P_0+\w_0)\sim \frac{\w_0^m}{\left[\w_0
+X+\imath \e\right]^n},
\ee
where $X$ is a combination of constants, $n=1,2$ and $m=0,1$, the
energy integral takes the form
\be
\int\dk{\w_0}\w_0^m\prod_{i=1}^{2+n}\frac{1}{\left[\w_0
+X_i+\imath\e\right]}=0.
\label{eq:loopcorr}
\ee
Clearly, this integral is a generalization of \eq{eq:tempint} and this
vanishes, just as for the loop corrections in the kernel of the \BS
equation from Ref.~\cite{Popovici:2010mb} and the diagram (b) from
above.  An identical calculation for the diagram (f), recalling that
the lower line corresponds to an antiquark propagator, leads us to the
fact that this integral is also vanishing.
\begin{figure}[t]
\vspace{0.5cm}
\centering\includegraphics[width=0.80\linewidth]{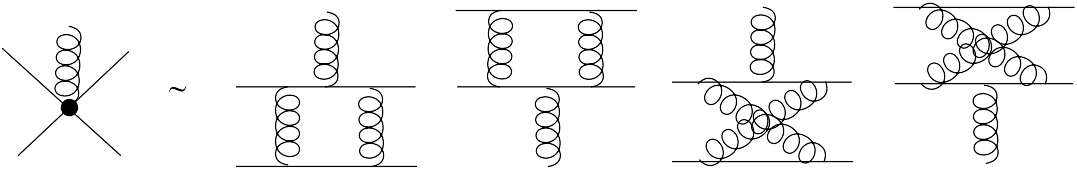}
\caption{\label{fig:diagr-e0}
Lowest order in the perturbative expansion of the proper four
quark-gluon vertex.  See text for details.}
\end{figure}
\begin{figure}[t]
\vspace{0.5cm}
\centering\includegraphics[width=0.7\linewidth]{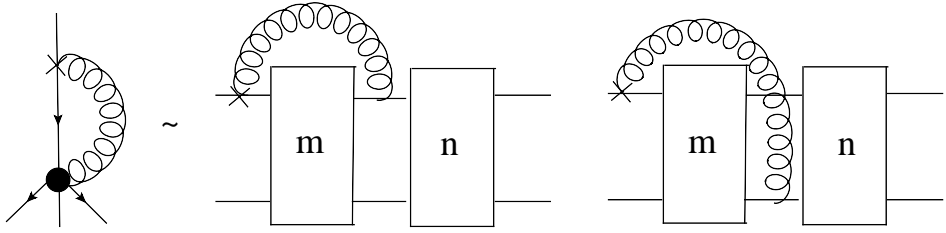}
\caption{\label{fig:diagr-e}
Perturbative expansion of the diagram (e). Boxes comprise $m$ and 
$n$ internal gluon lines, respectively, with $m, n\ge 1$.  See text
for details.}
\end{figure}

Finally, we are now in the position to show that the diagram (e),
containing the 4 quark-gluon vertex, is also vanishing.  The
argumentation is based on our previous findings, namely that the
diagrams (f) and (g), containing the 1PI quark Green's functions, are
zero.  Because we exclude the \YM vertices, the external gluon of the
five-point (proper) vertex can only couple to the (upper) quark or
(lower) antiquark line.  Then the 4 quark-gluon vertex at lowest order
contains the diagrams shown in \fig{fig:diagr-e0}. The crossed
diagrams vanish, based on integrals of the type \eq{eq:tempint} or
\eq{eq:loopcorr}, and in fact this can be generalized to the case with
an arbitrary number of external gluon legs. The remaining
contributions can then be included in the diagram (e) and can be
written as a combination of diagrams of the form shown in
\fig{fig:diagr-e}, where the boxes contain an arbitrary number of
internal gluon lines (ladder resummation).  On the other hand, as a
result of the \DS equation for the 1PI quark Green's functions (as in
\fig{fig:1PI_mr}), the 4-point functions contained in the diagrams (f)
and (g) can also be written as a ladder resummation, and with the
explicit gluon lines, diagrams (f), (g) coincide precisely with the
internal substructure of the two terms in the diagram
\fig{fig:diagr-e}.  Hence, the perturbative series of diagram (e) has
been reorganized such that although the function $\G_{\bar qq\bar
qq\si}$ itself does not vanish, this 5-point interaction vertex and
the gluon line on top of it \emph{do} indeed vanish at every order
perturbatively. In turn, this implies that our original assumption is
correct and the solution \eq{eq:sol1PI} is valid at every order in
perturbation theory. As a side remark, we note that the nonvanishing
of the 5-point function relates to the existence of three-quark bound
states in the Faddeev equation, in the ladder approximation
\cite{Popovici:2010ph}.

\subsection{Amputated Green's function }

In the following, we consider the \DS equation for the fully amputated
4-point quark-antiquark Green's function in the $s$-channel (which we
denote by $G^{(4)}$) and, with the simplifications outlined in the
previous Section, we will derive a solution to this equation.  The
motivation for studying the fully amputated Green's function in
addition to the proper function is the fact that the homogeneous \BS
equation is based on the amputated 4-point function.  Since the
reduction from \eq{eq:1pi_4quark} to \eq{eq:1pi_4quark1} is valid for
the 1PI Green's function, this must also hold for the amputated
Green's function.  We will explicitly verify that this is satisfied,
and will analyze the position of the poles and show that they relate
to the \BS equation for physical states.

\begin{figure}[t]
\vspace{0.5cm}
\centering\includegraphics[width=0.7\linewidth]{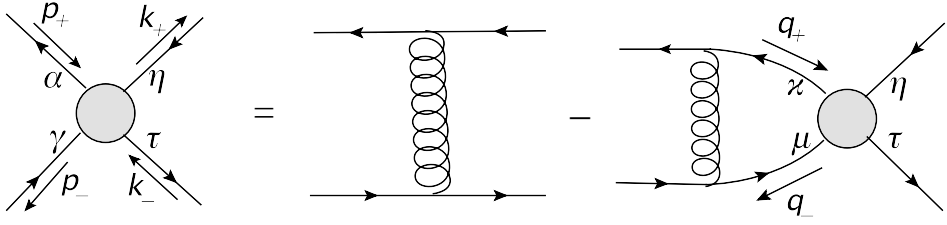}
\caption{\label{fig:amputated_mr}
Truncated Dyson-Schwinger equation for the fully amputated
quark-antiquark 4-point Green's function in the $s$-channel. See text
for details.}
\end{figure}

The appropriate \DS equation for the fully amputated 4-point
quark-antiquark Green's function can thus be obtained from the formula
\eq{eq:1pi_4quark1}, by replacing the 1PI Green's function $\G^{(4)}$
with the amputated Green's function $G^{(4)}$ using the expression
\eq{eq:connected_4quark} and cutting the legs.  This equation is the
inhomogeneous ladder \BS equation and writing it in terms of the usual
variables reads, in momentum space (see also \fig{fig:amputated_mr}):
\bea
 G^{(4)}_{\al\ga;\tau\eta}(p_+,p_-;k_+,k_-)&=& W_{\si\si}^{ab}(\vec
p-\vec k)\left[\G_{\bar qq\si}^{a}\right]_{\al\eta} \left[\G_{\bar q
q\si}^{b}\right]_{\tau\ga}
\nonumber\\
&&-\int\dk{q}\left[\G_{\bar qq\si}^{a(0)} W_{\bar
qq}(q_+)\right]_{\al\ka} \left[W_{\bar qq}(q_-)\G_{\bar
qq\si}^{b}\right]_{\mu \ga} W_{\si\si}^{ab}(\vec p-\vec q)
G^{(4)}_{\ka\mu;\tau\eta}(q_+,q_-;k_+,k_-).
\label{eq:4pointDS2}
\eea

In the above, the momenta of the quarks are given by $p_+=p+\xi P$,
$p_-=p-(1-\xi) P$ (similarly for $k$ and $q$), and $\xi$ is the
momentum sharing fraction.  $P$ indicates the dependence on the total
four momentum, which will become important for the investigation of
the bound-state contributions to the Green's function.  It comes as no
surprise that the inhomogeneous \BS equation is already in the ladder
truncation, similar to \eq{eq:1pi_4quark1}.

The right hand side of equation \eq{eq:4pointDS2} does not depend on
the external energy $p_0$, implying that the 4-point function
$G^{(4)}(p_+,p_-;k_+,k_-)$ has to be independent on the relative
energies $p_0, k_0$, and we further assume that $G^{(4)}$ depends on
the relative spatial momentum $\vec p-\vec k$.  We then identify the
antiquark line as before, replace the heavy quark and antiquark
propagators and vertices with the expressions
Eqs. (\ref{eq:quarkpropnonpert}, \ref{eq:antiquarkpropnonpert},
\ref{eq:feyn}, \ref{eq:antiqqsinp}), and perform the energy
integration. We arrive at the following expression:
\be
G^{(4)}_{\al\ga;\tau\eta}(P_0;\vec p-\vec k)= g^2
T^{a}_{\al\eta}T^{a}_{\tau \ga }W_{\si\si}(\vec p-\vec k)
+g^2T^{a}_{\al\ka}T^{a}_{\mu\ga}\frac{\imath}{P_0-2 {\cal
I}_r+2\imath\e} \int\dk{\vec q}W_{\si\si}(\vec p-\vec q)
G^{(4)}_{\ka\mu;\tau\eta}(P_0;\vec q-\vec k).
\label{eq:DS4point}
\ee

Having integrated out the energy, it is convenient to rewrite the
above formula back into coordinate space.  Using the definition
\eq{eq:g4fourier} and noting that the Fourier transform is as before
given via \eq{eq:convolution}, the equation \eq{eq:DS4point}
simplifies to (the separation $x$ corresponds to the momentum $\vec
p-\vec k$):
\be
G^{(4)}_{\al\ga;\tau\eta}(P_0;x) =g^2T^a_{\al\eta}T^{a}_{\tau
\ga}W_{\si\si}(x)+
g^2T^{a}_{\al\ka}T^{a}_{\mu\ga}\frac{\imath}{P_0-2{\cal
I}_r+2\imath\e}W_{\si\si}(x) G^{(4)}_{\ka\mu;\tau\eta}(P_0;x).
\ee

Again, we make a color decomposition of  the function $G^{(4)}$:
\be
G^{(4)}_{\al\ga;\tau\eta} =\de_{\al\ga}\de_{\tau
\eta}G^{(4)}_{1}+\de_{\al\eta}\de_{\tau \ga}G^{(4)}_{2},
\label{eq:g4color}
\ee
($G^{(4)}_{1}$ and $G^{(4)}_{2}$ are scalar functions), use the Fierz
identity, \eq{eq:fierz}, to sort out the color factors, and obtain the
following results for the components $G^{(4)}_1$, $G^{(4)}_2$:
\bea
G_1^{(4)}(P_0;x)&=&\left(\frac{g^2}{2}\right) \frac{(P_0-2{\cal
I}_r)^2 W_{\si\si}(x) } {[P_0-2{\cal I}_r+\imath
\frac{g^2}{2N_c}W_{\si\si}(x)+2\imath\e] [P_0-2{\cal I}_r-\imath
\frac{g^2}{2}\left(N_c-\frac 1N_c\right)W_{\si\si}(x) +2\imath\e]},
\\
G_2^{(4)}(P_0; x)&=&-\left(\frac{g^2}{2N_c}\right) \frac{(P_0-2{\cal
I}_r) W_{\si\si}(x) } {P_0-2{\cal I}_r+\imath
\frac{g^2}{2N_c}W_{\si\si}(x) +2\imath\e}.
\eea
Replacing these results in the formula \eq{eq:g4color}, we get the
final result for the function $G^{(4)}$:
\be
G^{(4)}_{\al\ga;\tau\eta}(P_0;x)=\frac{g^2}{2} \frac{\left(P_0-2{\cal
I}_r\right) W_{\si\si}(x)}{P_0-2{\cal I}_r+\imath
\frac{g^2}{2N_c}W_{\si\si}(x) +2\imath\e} \left[
\de_{\al\ga}\de_{\tau\eta} \frac{(P_0-2{\cal I}_r) } {P_0-2{\cal
I}_r-\imath g^2C_F W_{\si\si}(x) +2\imath\e}-
\de_{\al\eta}\de_{\tau\ga}\frac{1}{N_c} \right].
\label{eq:4pointfinal}
\ee
A direct calculation shows that our result for the amputated 4-point
function is related to the result \eq{eq:sol1PI} for the 1PI Green's
function, via the formula \eq{eq:connected_4quark} in the $s$-channel
(or alternatively, \fig{fig:1PI_amputated}):
\bea
\G^{(4)}_{\al\ga\tau\eta}(p_+,p_-,k_-,k_+)=
G^{(4)}_{\al\ga;\tau\eta}(p_+,p_-;k_+,k_-)
-W_{\si\si}^{ab}(\vec p-\vec k)\left[\G_{\bar
qq\si}^{a}\right]_{\al\eta} \left[\G_{\bar q
q\si}^{b}\right]_{\tau\ga}.
\label{eq:rel}
\eea

A few comments regarding the structure of the above equation are in
order. Notice that despite the truncation, the denominator structure
of the 1PI and amputated Green's functions is identical in both color
channels. Moreover, in this approach the physical and nonphysical
poles disentangle automatically.  Using the form \eq{eq:Wsisi} for the
temporal gluon propagator, the denominator factor of the color singlet
channel can be rewritten in the form
\be
P_0-g^2\int_{r}\frac{\dk{\vec{\w}}D_{\si\si}(\vec{\w})}{\vec{\w}^2}
C_F\left[1-e^{\imath\vec{\w}\cdot\vec{x}}\right]
\ee
and hence the bound state (infrared confining) energy $P_{0\;
res}(x)=\si|\vec x|$ emerges as the finite pole position of the
resonant component (first term in the bracket of \eq{eq:4pointfinal}
or \eq{eq:sol1PI}), for arbitrary number of colors, assuming that
$D_{\si\si}(\vec\w^2)\sim 1/\vec\w^2$ as indicated on the lattice
\cite{Quandt:2008zj}. This provides an explicit analytical dependence
of the 4-point Green's function on the $\bar qq$ bound state energy,
which results from the homogeneous \BS equation from
Ref.~\cite{Popovici:2010mb}. The overall denominator factor for
\eq{eq:4pointfinal} is part of the normalization and has the explicit
form
\be
P_0-g^2\int_{r}\frac{\dk{\vec{\w}}D_{\si\si}(\vec{\w})}{\vec{\w}^2}
\left[C_F+\frac{1}{2N_c}e^{\imath\vec{\w}\cdot\vec{x}}\right]
\ee
which (like the quark propagator) implies a pole position at infinity
when the regularization is removed. This factor does not appear in the
homogeneous \BS equation. Notice that the result, \eq{eq:4pointfinal},
is independent of $\xi$ (the momentum sharing fraction). That the
physical pole position should be independent of $\xi$ is clear (c.f.
Refs.~\cite{Popovici:2010mb,Alkofer:2002bp}); that the rest is
independent of $\xi$ is presumably a consequence of the truncation
scheme considered here.

\section{4-point Green's functions for diquarks}

\begin{figure}
\vspace{0.5cm}
\centering\includegraphics[width=0.9\linewidth]{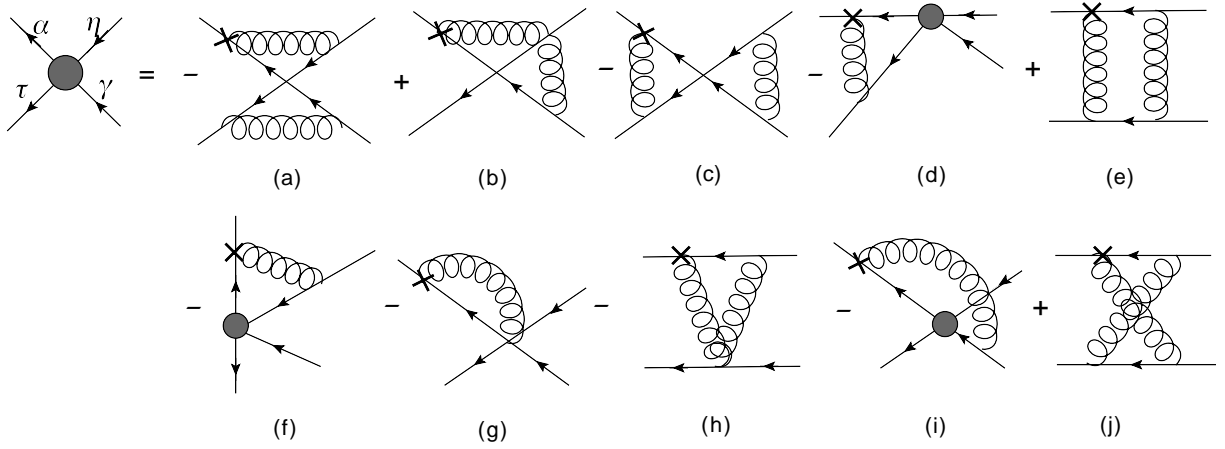}
\caption{\label{fig:1PIdiquarks}
Diagrammatic representation of the \DS equation for the 1PI 4-point
diquark Green's function. Same conventions as in \fig{fig:1PI} apply.}
\end{figure}
Let us now consider the diquark 4-point Green's function. This can be
easily obtained from the equation \eq{eq:1pi_4quark} for
quark-antiquark systems, by interchanging the quark legs and inserting
the appropriate minus signs. We obtain (see also the diagrammatic
representation from \fig{fig:1PIdiquarks}):
\bea
\lefteqn{ \ev{\imath\ov{q}_{\al x}\imath\ov{q}_{\tau w}\imath q_{\ga
z}\imath q_{\eta t}} = [g\ga^0T^a]_{\al\ba}\int dy\, \de(x-y)
}\nonumber\\
&&\times\left\{ 
-\left[
\ev{\imath\ov\chi_{\ba x}\imath\chi_{\ka}}
\ev{\imath\ov{q}_{\ka}\imath q_{\ga z}\imath \si^{c}_{\la}}
\ev{\imath\rho^c_{\la}\imath\rho^d_{\nu}}
\right]
\left[
\ev{\imath\ov{q}_{\tau w}\imath q_{\mu}\imath \si^{d}_{\nu}}
\ev{\imath\ov\chi_{\mu}\imath\chi_{\de}}
\ev{\imath\ov{q}_{\de}\imath q_{\eta t}\imath \si^{b}_{\e}}
\ev{\imath\rho^b_{\e}\imath\rho^a_y}
\right]
\right.\nonumber\\
&&
+\left[
\ev{\imath\ov\chi_{\ba x}\imath\chi_{\ka}}
\ev{\imath\ov{q}_{\ka}\imath q_{\ga z}\imath \si^{c}_{\la}}
\ev{\imath\rho^c_{\la}\imath\rho^d_{\nu}}
\ev{\ov{q}_{\tau w}\imath q_{\eta t}\imath \si^{d}_{\nu}\imath
 \si^{b}_{\mu}}
\ev{\imath\rho^b_{\mu}\imath\rho^a_y}
\right]
\nonumber\\
&&-
\left[
\ev{\imath\ov\chi_{\ba x}\imath\chi_{\ka}}
\ev{\imath\ov{q}_{\ka}\imath q_{\ga z}\imath \si^{c}_{\la}}
\ev{\imath\rho^c_{\la}\imath\rho^d_{\de}}
\right]
\left[
\ev{\imath\ov{q}_{\tau w}\imath q_{\nu}\imath \si^{b}_{\mu}}
\ev{\imath\ov\chi_{\nu}\imath\chi_{\e}}
\ev{\imath\ov{q}_{\e}\imath q_{\eta t}\imath \si^{d}_{\de}}
\ev{\imath\rho^b_{\mu}\imath\rho^a_y}
\right]\nonumber\\
&&-
\left[
\ev{\imath\ov\chi_{\ba x}\imath\chi_{\ka}}
\ev{\imath\ov{q}_{\ka}\imath\ov{q}_{\la}\imath q_{\ga z}\imath 
q_{\eta t}}
\ev{\imath\ov{q}_{\tau w}\imath q_{\nu}\imath \si^{b}_{\mu}}
\ev{\imath\ov\chi_{\nu}\imath\chi_{\la}}
\ev{\imath\rho^b_{\mu}\imath\rho^a_y}
\right]\nonumber\\
&&+
\left[
\ev{\imath\ov\chi_{\ba x}\imath\chi_{\de}}
\ev{\imath\ov{q}_{\de}\imath q_{\eta t}\imath \si^{c}_{\e}}
\ev{\imath\rho^c_{\e}\imath\rho^d_{\ka}}
\right]
\left[
\ev{\imath\ov{q}_{\tau w}\imath q_{\nu}\imath \si^{b}_{\mu}}
\ev{\imath\ov\chi_{\nu}\imath\chi_{\la}}
\ev{\imath\ov{q}_{\la}\imath q_{\ga z}\imath \si^{d}_{\ka}}
\ev{\imath\rho^b_{\mu}\imath\rho^a_y}
\right]\nonumber\\
&&-
\left[
\ev{\imath\ov\chi_{\ba x}\imath\chi_{\ka}}
\ev{\imath\ov{q}_{\ka}\imath\ov{q}_{\tau w}
\imath q_{\ga z}\imath q_{\la}}
\ev{\imath\ov\chi_{\la}\imath\chi_{\de}}
\ev{\imath\ov{q}_{\de}\imath q_{\eta t}\imath \si^{b}_{\e}}
\ev{\imath\rho^b_{\e}\imath\rho^a_y}
\right]\nonumber\\
&&+
\left[
\ev{\imath\ov\chi_{\ba x}\imath\chi_{\ka}}
\ev{\imath\ov{q}_{\ka}\imath\ov{q}_{\tau w}\imath q_{\ga z}
\imath q_{\eta t}\si^{b}_{\la}}
\ev{\imath\rho^b_{\la}\imath\rho^a_y}
\right]\nonumber\\
&&-
\left[
\ev{\imath\ov\chi_{\ba x}\imath\chi_{\de}}
\ev{\imath\ov{q}_{\de}\imath q_{\eta t}\imath \si^{c}_{\e}}
\ev{\imath\rho^c_{\e}\imath\rho^d_{\ka}}\right]
\left[
\ev{\imath\ov{q}_{\tau w}\imath q_{\ga z}\imath\si^d_{\ka}
\imath\si^b_{\la}}
\ev{\imath\rho^b_{\la}\imath\rho^a_y}
\right]\nonumber\\
&&-
\left[
\ev{\imath\ov\chi_{\ba x}\imath\chi_{\nu}}
\ev{\imath\ov{q}_{\nu}\imath\ov{q}_{\tau w}\imath q_{\mu}
\imath q_{\eta t}}
\ev{\imath\ov\chi_{\mu}\imath\chi_{\ka}}
\ev{\imath\ov{q}_{\ka}\imath q_{\ga z}\imath \si^{b}_{\la}}
\ev{\imath\rho^b_{\la}\imath\rho^a_y}
\right]\nonumber\\
&&+
\left.
\left[
\ev{\imath\ov\chi_{\ba x}\imath\chi_{\de}}
\ev{\imath\ov{q}_{\de}\imath q_{\eta t}\imath \si^{c}_{\e}}
\ev{\imath\rho^c_{\e}\imath\rho^d_{\nu}}
\right]
\left[
\ev{\imath\ov{q}_{\tau w}\imath q_{\mu}\imath \si^{d}_{\nu}}
\ev{\imath\ov\chi_{\mu}\imath\chi_{\ka}}
\ev{\imath\ov{q}_{\ka}\imath q_{\ga z}\imath \si^{b}_{\la}}
\ev{\imath\rho^b_{\la}\imath\rho^a_y}
\right]
\right\}+\dots . 
\label{eq:1pi_4diquark}
\eea
As in the $\bar qq$ case, the dots represent the $\vec A$ vertex terms
which are not considered here and the tree-level temporal quark-gluon
vertex has been replaced with its expression
\eq{eq:treelevelquarkvertex1}.  Also, notice that the diquark is
antisymmetric under the exchange of two quark legs
($\al\leftrightarrow\tau$ or $\ga\leftrightarrow\eta$).  We must
therefore explicitly keep track of which quark is which and thus
introduce flavor structure with an arbitrary number of equal mass
flavors (before, we had to distinguish between quark and antiquark).

\begin{figure}[t]
\vspace{0.5cm}
\centering\includegraphics[width=0.23\linewidth]{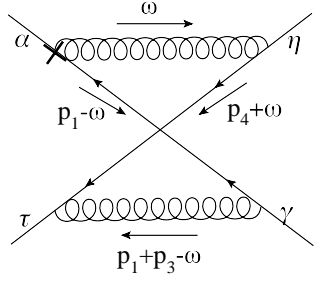}
\caption{\label{fig:diagr-a-diq}
Crossed ladder diagram contributing to the 1PI 4-point Green's
function in the diquark channel.}
\end{figure}

As before, we analyze the diagrammatic representation from
\fig{fig:1PIdiquarks} and show that the same type of cancellations
occur. Starting with the diagram (a) in momentum space, we notice that
this is a crossed ladder type exchange diagram (see also
\fig{fig:diagr-a-diq}):
\bea
\lefteqn{ \int\dk{\w}\left[\G_{\bar qq\si
\al\ba}^{(0)a}(p_1,\w-p_1,-\w)W_{\bar qq \ba\ka}(p_1-\w) \G_{\bar
qq\si \ka\ga}^{d}(p_1-\w,p_3,\w-p_1-p_3)\right]}\nonumber\\
&&\times\left[\G_{\bar q q\si\tau\mu}^{c}(p_2,p_4+\w,p_1+p_3-\w)
W_{\bar qq\mu\de}(-p_4-\w) \G_{\bar
qq\si\de\eta}^{b}(-p_4-\w,p_4,\w)\right] W_{\si\si}^{ab}(\w)
W_{\si\si}^{dc}(p_1+ p_3-\w).
\label{eq:diagr-a-diq}
\eea
It has been already shown that the integral over the quark propagators
(with the same Feynman prescription) vanishes and thus the diagram (a)
is zero.  A similar type of integral arises in the diagram (j), and
hence this term is also not giving a contribution. Further, the
diagrams (b) and (h) are zero, since the corresponding quark-2 gluon
vertex vanishes according to the \ST identity, \eq{eq:stid5pt}.  As in
the case of the $\bar qq$ system, let us for the moment assume that
the integrals (f), (i), containing the diquark 4-point vertex, and
(g), containing the 5-point function $\G_{\bar q\bar q qq\si}$ are
also zero, and solve the \DS equation with the remaining terms. Having
derived the solution, we will then return to the diagrams (f),(i) and
(g) and show that they vanish.  The \DS equation for the diquark
4-point function $\G^{(4d)}$ with the remaining diagrams,
i.e. diagrams (c), (d) and (e) explicitly reads, in momentum space:
\bea
\lefteqn{\G_{\al\tau\ga\eta}^{(4d)}(p_1,p_2,p_3,p_4)=
-\int\dk{\w}\left[\G_{\bar qq\si \al\ba}^{(0)a}(p_1,-\w-p_1,\w)W_{\bar
qq \ba\ka}(p_1+\w) \G_{\bar qq\si
\ka\ga}^{c}(p_1+\w,p_3,-\w-p_1-p_3)\right]}\nonumber\\
&&\times\left[\G_{\bar qq\si\tau\nu}^{b}(p_2,\w-p_2,-\w) W_{\bar
qq\nu\e}(p_2-\w) \G_{\bar
qq\si\e\eta}^{d}(p_2-\w,p_4,p_1+p_3+\w)\right] W_{\si\si}^{ab}(-\w)
W_{\si\si}^{cd}(p_1+ p_3+\w) \nonumber\\
&&+\int\dk{\w}\left[\G_{\bar qq\si
\al\ba}^{(0)a}(p_1,-\w-p_1,\w)W_{\bar qq \ba\de}(p_1+\w) \G_{\bar
qq\si \de\eta}^{d}(p_1+\w,p_4,-\w-p_1-p_4)\right]\nonumber\\
&&\times\left[\G_{\bar qq\si\tau\nu}^{b}(p_2,\w-p_2,-\w) W_{\bar
qq\nu\la}(p_2-\w) \G_{\bar
qq\si\la\ga}^{c}(p_2-\w,p_3,p_1+p_4+\w)\right] W_{\si\si}^{ab}(-\w)
W_{\si\si}^{dc}(p_1+ p_4+\w) \nonumber\\
&&-\int\dk{\w}\left[\G_{\bar qq\si
\al\ba}^{(0)a}(p_1,-\w-p_1,\w)W_{\bar qq \ba\la}(p_1+\w)\right]
\left[\G_{\bar q q\si\tau\nu}^{b}(p_2,\w-p_2,-\w) W_{\bar
qq\nu\ka}(p_2-\w)\right] \nonumber\\
&&\times
\G_{\la\ka\ga\eta}^{(4d)}(p_1+\w,p_2-\w,p_3,p_4)W_{\si\si}^{ab}(-\w).
\label{eq:1PImomdiq}
\eea
Inserting the expressions for the quark propagator
\eq{eq:quarkpropnonpert} and vertex \eq{eq:feyn}, resolving the color
factors using the Fierz identity, \eq{eq:fierz}, and simplifying the
energy integrals, we have (noting the flavor channels with the $\de^f$
factors):
\bea
\G^{(4d)}_{\al\tau\ga\eta}(p_1,p_2,p_3,p_4) &=&
\frac{1}{ P^0-2m-2 {\cal I}_{r} +2\imath\e}\nonumber\\
&&\left\{
-\imath\left(\frac{g^2}{2N_c}\right)^2
\de_{\al\ga}^f\de_{\tau\eta}^f
\left[\left(N_c^2+1\right)\de_{\al\ga}\de_{\tau\eta}
-2N_c\de_{\al\eta}\de_{\tau \ga}\right]
\int\dk{\vec\w}W_{\si\si}(\vec\w)
W_{\si\si}(\vec p_1+\vec p_3+\vec\w)\right.\nonumber\\
&&+\imath\left(\frac{g^2}{2N_c}\right)^2
\de_{\al\eta}^f\de_{\tau \ga }^f
\left[\left(N_c^2+1\right)\de_{\al\eta}\de_{\tau \ga}
-2N_c\de_{\al\ga}\de_{\tau\eta}\right]
\int\dk{\vec\w}W_{\si\si}(\vec\w)
W_{\si\si}(\vec p_1+\vec p_4+\vec\w)\nonumber\\
&&+
\frac{g^2}{2N_c} \de_{\al\la}^f\de_{\tau\ka}^f
\left[N_c\de_{\al\ka}\de_{\tau\la}
-\de_{\al\la}\de_{\tau\ka}\right]
\int\dk{\w_0} \left[
\frac{1}{\w_0+ p_1^0-m- {\cal I}_{r} +\imath\e}
-\frac{1}{\w_0- p_2^0+m+ {\cal I}_{r} -\imath\e}
\right]\nonumber\\
&&\times\left.\int\dk{\vec\w}W_{\si\si}(\vec\w)
\G^{(4d)}_{\la\ka\ga\eta}(p_1+\w,p_2-\w,p_3,p_4)\right\}.
\label{eq:1PIsimpl-di}
\eea

Noticing the distinctive energy, momentum and flavor structure of the
equation, a suitable Ansatz for the diquark function is
\be
\G^{(4)}_{\al\tau\ga\eta} (p_1,p_2,p_3,p_4)
=\de_{\al\ga}^{f}\de_{\tau\eta}^{f}\G^{(c)}_{\al\tau\ga\eta}(P_0;
\vec p_1+\vec p_3)
+\de_{\al\eta}^{f}\de_{\tau\ga}^{f}\G^{(p)}_{\al\tau\ga\eta}(P_0;
\vec p_1+\vec p_4)
\label{eq:g4flavour1PIdi}
\ee
where the superscripts $c$ and $p$ stand for the crossed and parallel
configurations, respectively, $P_0=p_1^0+p_2^0$ as before, and
$\G^{(c,p)}_{\al\tau\ga\eta}$ still retain the color indices.  With
these notations, \eq{eq:1PIsimpl-di} decouples via the flavor
structure to:
\bea
\G^{(c)}_{\al\tau\ga\eta}(P_0;\vec p_1+\vec p_3)
&=&(-\imath)\frac{g^2}{2N_c} \frac{1}{ P^0-2m-2 {\cal I}_{r}
+2\imath\e}\nonumber\\
&&\times\left\{
\frac{g^2}{2N_c}\left[\left(N_c^2+1\right)\de_{\al\ga}\de_{\tau\eta}
-2N_c\de_{\al\eta}\de_{\tau \ga}\right]
\int\dk{\vec\w}W_{\si\si}(\vec\w) W_{\si\si}(\vec p_1+\vec
p_3+\vec\w)\right.\nonumber\\
&&+\left.  \left[N_c\de_{\al\ka}\de_{\tau
\la}-\de_{\al\la}\de_{\tau\ka}\right]
\int\dk{\vec\w}W_{\si\si}(\vec\w) \G^{(c)}_{\la\ka\ga\eta}(P_0;\vec
p_1+\vec p_3+\vec\w) \right\},
\label{eq:1PIsimpl-di1}\\
\G^{(p)}_{\al\tau\ga\eta}(P_0;\vec p_1+\vec p_4)
&=&\imath\frac{g^2}{2N_c} \frac{1}{ P^0-2m-2 {\cal I}_{r}
+2\imath\e}\nonumber\\
&&\times\left\{
\frac{g^2}{2N_c}\left[\left(N_c^2+1\right)\de_{\al\eta}\de_{\tau\ga}
-2N_c\de_{\al\ga}\de_{\tau\eta}\right]
\int\dk{\vec\w}W_{\si\si}(\vec\w) W_{\si\si}(\vec p_1+\vec
p_4+\vec\w)\right.\nonumber\\
&&-\left.  \left[N_c\de_{\al\ka}\de_{\tau\la}
-\de_{\al\la}\de_{\tau\ka}\right]
\int\dk{\vec\w}W_{\si\si}(\vec\w) \G^{(p)}_{\la\ka\ga\eta}(P_0;\vec
p_1+\vec p_4+\vec\w) \right\}.
\label{eq:1PIsimpl-di2}
\eea

Fourier transforming to configuration space as in the previous section
and with $x, y$ representing the separations associated with $\vec
p_1+\vec p_4$, $\vec p_1+\vec p_3$, respectively, the equations read
\bea
\G^{(c)}_{\al\tau\ga\eta}(P_0;y) &=&(-\imath)\frac{g^2}{2N_c}
\frac{1}{ P^0-2m-2 {\cal I}_{r} +2\imath\e} \left\{ \frac{g^2}{2N_c}
\left[(N_c^2+1) \de_{\al\ga}\de_{\tau\eta}-2N_c
\de_{\al\eta}\de_{\tau\ga}\right] W_{\si\si}(y)^2\right.\nonumber\\
&&\left.+\left[N_c\de_{\al\ka}\de_{\tau\la}
-\de_{\al\la}\de_{\tau\ka}\right]
W_{\si\si}(y)\G^{(c)}_{\la\ka\ga\eta}(P_0;y)\right\},\\
\G^{(p)}_{\al\tau\ga\eta}(P_0;x) &=&(+\imath)\frac{g^2}{2N_c}
\frac{1}{ P^0-2m-2 {\cal I}_{r} +2\imath\e} \left\{ \frac{g^2}{2N_c}
\left[(N_c^2+1) \de_{\al\eta}\de_{\tau\ga}-2N_c
\de_{\al\ga}\de_{\tau\eta}\right] W_{\si\si}(x)^2\right.\nonumber\\
&&\left.-\left[N_c\de_{\al\ka}\de_{\tau\la}
-\de_{\al\la}\de_{\tau\ka}\right]
W_{\si\si}(x)\G^{(p)}_{\la\ka\ga\eta}(P_0;x)\right\}.
\eea
Further, we make the color decomposition:
\be
\G^{(c,p)}_{\al\tau\ga\eta}
=\de_{\al\ga}\de_{\tau\eta}\G^{(c,p;1)}
+\de_{\al\eta}\de_{\tau\ga}\G^{(c,p;2)},
\label{eq:g4color1PIdi}
\ee
where $\G^{(c,p;1,2)}$ are now scalar dressing functions. Defining the
functions
\bea
f_{\pm}(P_0;x)=(-\imath) \left(\frac{g^2}{2N_c}\right)^2
\frac{\left(N_c\mp 1\right)^2W_{\si\si}(x)^2} { P_0-2m-2 {\cal I}_{r}
-\imath\frac{g^2}{2N_c}(1\mp N_c) W_{\si\si}(x)+2\imath\e}
\eea
it is then straightforward to show that the solution to the above
equations can be written
\begin{subequations}
\bea
2\G^{(c1)}(P_0;y)&=&f_{+}(P_0;y)+f_{-}(P_0;y),\\
2\G^{(c2)}(P_0;y)&=&f_{+}(P_0;y)-f_{-}(P_0;y),\\
2\G^{(p1)}(P_0;x)&=&f_{-}(P_0;x)-f_{+}(P_0;x)=-2\G^{(c2)}(P_0;x),\\
2\G^{(p2)}(P_0;x)&=&-f_{+}(P_0;x)-f_{-}(P_0;x)=-2\G^{(c1)}(P_0;x).
\eea
\end{subequations}
Putting the components together, we find
\bea
\G_{\al\tau\ga\eta}(P_0;x,y)\!&=&\!
\de_{\al\ga}^f\de_{\tau\eta}^f
\left[\de_{\al\ga}\de_{\tau\eta}\G^{(c1)}(P_0;y)+
\de_{\al\eta}\de_{\tau \ga}\G^{(c2)}(P_0;y)\right]+
\de_{\al\eta}^f\de_{\tau\ga}^f
\left[\de_{\al\ga}\de_{\tau\eta}\G^{(p1)}(P_0;x)+
\de_{\al\eta}\de_{\tau \ga}\G^{(p2)}(P_0;x)\right]\nonumber\\
&=&\frac{1}{2}\left\{
\left(\de_{\al\ga}\de_{\tau\eta}+\de_{\al\eta}\de_{\tau \ga}\right)
\left[ \de_{\al\ga}^f\de_{\tau\eta}^f  f_{+}(P_0;y)- 
\de_{\al\eta}^f\de_{\tau\ga}^f f_{+}(P_0;x)\right]\right. \nonumber\\
&&+\left.
\right(\de_{\al\ga}\de_{\tau\eta}-\de_{\al\eta}\de_{\tau \ga}\left)
\left[ \de_{\al\ga}^f\de_{\tau\eta}^f  f_{-}(P_0;y)
+ \de_{\al\eta}^f\de_{\tau\ga}^f f_{-}(P_0;x)\right]
\right\}.
\label{eq:diqsolution1}
\eea

As in the case of the $\bar qq$ systems, with this solution we return
to the diagrams (f), (i) and (g). Writing out the explicit form of the
energy integrals we notice that their form is identical to the
quark-antiquark case, since the $\e$ prescription is similar,
regardless of the internal quark (or antiquark) propagator. Thus,
these diagrams are also vanishing.

Analyzing the pole structure of the above solution,
\eq{eq:diqsolution1}, we notice that, as in the case of the $\bar qq$
systems, we have two different pole conditions (for either separation
$x$ or $y$):
\be
P_{0\; res}(x)-2m-2 {\cal I}_{r}-\imath\frac{g^2}{2N_c}(1\pm N_c)
W_{\si\si}( x)=0
\ee
which can be rewritten as 
\be
P_{0\;res}(x)=2m+g^2\int_{r}\frac{\dk{\vec{\w}}
D_{\si\si}(\vec{\w})}{\vec{\w}^2}
\left[C_F- \frac{1\pm N_c}{2N_c}
e^{\imath\vec{\w}\cdot\vec{x}}\right].
\ee

With the plus sign (corresponding to the color antisymmetric term or
$f_{-}$ in \eq{eq:diqsolution1} above), we find the condition for the
finite pole position (i.e., finite integral even with infrared
enhanced $D_{\si\si}$) $P_{0\; res}(x)$, $N_c=2,-1$; but with the
minus sign ($f_{+}$, color symmetric), we obtain $N_c=-2,1$. Hence,
the only physical solution, with $N_c=2$, corresponds to a color
antisymmetric and flavor symmetric configuration, in agreement with
our findings from Ref.~\cite{Popovici:2010mb} that bound states exist
only for color antisymmetric $SU(2)$ baryons (in that case flavor
symmetry was implicit).  Indeed, the physical poles in the flavor
channels are the same (both correspond to $f_{-}$), as demanded by the
symmetry of the system.  The homogeneous equation makes no reference
to the flavor structure and can only have the flavor symmetric part.
Also notice that whereas for the $\bar qq$ system there was a common
unphysical pole in the normalization of $\G^{(4)}$ or $G^{(4)}$ and
then a distinct physical pole for the color singlet channel, in the
diquark case each channel has a separate pole.

\section{Summary and conclusions}

In this paper, the \DS equations for the 1PI and amputated 4-point
Green's functions, for quark-antiquark and diquark systems, have been
considered. At leading order in the heavy quark mass expansion and
with the truncation to include only the (nonperturbative) gluon
propagator and neglect the pure \YM vertices and higher order
interactions, analytic solutions for the Green's functions have been
obtained.  We have found that the physical and unphysical poles
disentangle, in contrast to the conventional phenomenological
calculations. Indeed, this is one of the main results of this work,
since in contemporary studies the interpretation of the spurious
states appearing in the \BS equation is still under debate.  These
results may hopefully be of further use in the construction of
phenomenological models of mesons and baryons.

The color singlet pole of the 4-point Green's function for $\bar qq$
systems leads directly to the linearly rising potential (confining
energy for the bound state), for arbitrary number of colors, and in
turn this can be related to the $\bar qq$ bound state energy arising
from the homogeneous \BS equation (considered in
Ref.~\cite{Popovici:2010mb}).  The second pole, which is situated at
infinity as the (infrared) regularization is removed, is nonphysical
and is part of the normalization; it is similar to the poles appearing
in the gap equation and in the Faddeev vertex for baryons
\cite{Popovici:2010ph}. This simply means that the only relevant
quantity is the bound state energy, i.e.  the pole of the resonant
component.  In the case of the diquarks, the only physical solution,
for $N_c=2$ colors, corresponding to a color antisymmetric and flavor
symmetric configuration, again verifies our result from
Ref.~\cite{Popovici:2010mb} that bound states only appear for $SU(2)$
baryons and otherwise the system is not physically allowed.  The
second pole, corresponding to a flavor antisymmetric configuration, is
not contained within the homogeneous \BS equation (there, the flavor
symmetry is implicit). We thus confirm that within the scheme studied
here, the homogeneous \BS equation furnishes exactly the right
physical poles with no other solution and these are explicitly
contained within the full Green's functions.

\begin{acknowledgments}
C.P. has been supported by the Deutscher Akademischer Austausch Dienst
(DAAD) and partially by the EU-RTN Programme, Contract
No.MRTN--CT-2006-035482, \lq\lq Flavianet''. P.W. and H.R. have been
supported by the Deutsche Forschungsgemeinschaft (DFG) under contracts
no. DFG-Re856/6-2,3.
\end{acknowledgments}

\appendix*

\section {Slavnov-Taylor identities}

In this Appendix we consider the \ST identity for the quark-2 gluon
vertex appearing in the text.  The derivation is similar to the \ST
identity for the quark-gluon vertex presented in
Ref.~\cite{Popovici:2010mb}.  We start with the full QCD action in the
standard, second order formalism:
\bea
{\cal S}_{QCD}=\int d^4x\left\{
\ov{q}_\al\left[\imath\ga^0\partial_{0}+gT^a\ga^0\si^a+\imath\s{
\vec{\ga}}{\div}-gT^a\s{\vec{\ga}}{\vec{A}^a}-m\right]_{\al\ba}q_
\ba-\frac{1}{4}F_{\mu\nu}^aF^{a\mu\nu} \right\}
\label{eq:qcdaction}
\eea
where the (antisymmetric) field strength tensor is defined in terms of
the gauge potential $A_{\mu}^a$:
\be
F_{\mu\nu}^a=\partial_{\mu}A_{\nu}^a-\partial_{\nu}A_{\mu}^a+gf^
{abc}A_{\mu}^bA_{\nu}^c.
\ee
The action is invariant under a local $SU(N_c)$ gauge transform
characterized by the parameter $\th_x^a$:
\be
U_x=\exp{\left\{-\imath\th_x^aT^a\right\}}
\ee
such that for infinitesimal $\th_x^a$, the fields transform as (with
the notation $\si\equiv A_0$)
\bea
&&\de \si_a=-\frac{1}{g}\partial_0\th^a-f^{abc}\si^b\th^c,\,\,\,
 \de\vec{A}^a=\frac{1}{g}\div\th^a-f^ {abc}\vec{A}^b\th^c,
\nonumber\\
&&\de q_\al=-\imath\th^a\left[T^a\right]_{\al\ba}q_\ba,\,\,\,
\de\ov{q}_\al=\imath\th^a\ov{q}_\ba
\left[T^a\right]_{\ba\al}.
\eea
The action \eq{eq:qcdaction} is invariant under gauge transformations
and hence the functional integral $Z$ is divergent by virtue of a zero
mode. To surmount this problem we use the Faddeev-Popov technique and
introduce a gauge-fixing term along with an associated ghost term.  In
Coulomb gauge ($\s{\div}{\vec{A}^a}=0$), the new term in the action
reads:
\be
{\cal S}_{FP}=\int d^4x\left[-\la^a\s{\div}{\vec{A}^a}-\ov{c}^a\s{
\div}{\vec{D}^{ab}}c^b\right],
\ee
where the gauge fixing condition is implemented with the help of the
Lagrange multiplier $\la^a$, and $\ov{c}^a$ and $c^b$ are the
Grassmann-valued ghost fields.  The spatial covariant derivative (in
the adjoint representation) is given by:
\be
\vec{D}^{ab}=\de^{ab}\div-gf^{acb}\vec{A}^c.
\ee

The action is invariant under a Gauss-BRST transform
\cite{Zwanziger:1998ez} whereby the infinitesimal spacetime-dependent
parameter $\th_x^a$ is factorized into two Grassmann-valued
components: $\th_x^a=c_x^a\de\la_t$. In Coulomb gauge, the gauge
fixing term does not involve any time derivatives and hence one can
define the infinitesimal variation $\de\la_t$ (not to be confused with
the Lagrange multiplier $\la^a$) to be \emph{time-dependent}.  The
variations of the new fields read:
\be
\de\ov{c}^a=\frac{1}{g}\la^a\de\la_t,\;\;\;\;\de c^a=-\frac{1}{2}
f^{abc}c^bc^c\de\la_t,\;\;\;\;\de\la^a=0.
\ee
The generating functional is given by
\be
Z[J]=\int{\cal D}\Phi\exp{\left\{\imath{\cal S}_{QCD}+
\imath{\cal S}_{FP}+\imath{\cal S}_s\right\}},
\ee
with the source term
\be
{\cal S}_s=\int d^4x\left[\ro^a\si^a+\s{\vec{J}^a}{\vec{A}^a}+
\ov{c}^a\et^a+\ov{\et}^ac^a+\xi^a\la^a+\ov{q}_\al\chi_\al+\ov{\chi}
_\al q_\al\right].
\ee

The \ST identities in Coulomb gauge are derived from the observation
that the Gauss-BRST transform can be regarded as a change of
integration variables under which the generating functional is
invariant. Provided that the Jacobian factor is trivial
\cite{Watson:2006yq}, we are left with an equation where only the
source term varies:
\bea
0&=&\left.\int{\cal D}\Phi\frac{\de}{\de\left[\imath\de\la_t\right]
}\exp{\left\{\imath{\cal S}_{QCD}+\imath{\cal S}_{FP}+ \imath{\cal
S}_s+\imath\de{\cal S}_s\right\}}\right|_{\de\la_t=0}
\nonumber\\
&=&\int{\cal D}\Phi\exp{\left\{\imath{\cal S}_{QCD}+\imath{\cal S}
_{FP}+\imath{\cal S}_s\right\}}\int d^4x\de(t-x_0)\left\{
-\frac{1}{g}\left(\partial_x^0\ro_x^a\right)c_x^a+f^{abc}\ro_x^a
\si_x^bc_x^c
\right.\nonumber\\&&\left.
-\frac{1}{g}J_{ix}^a\nabla_{ix}c_x^a+f^{abc}J_{ix}^aA_{ix}^bc_x^c-
\imath\ov{\chi}_{\al x}c_x^aT_{\al\ba}^aq_{\ba x}-\imath c_x^a
\ov{q}_{\ba x}T_{\ba\al}^a\chi_{\al x}+\frac{1}{g}\la_x^a\et_x^a+
\frac{1}{2}f^{abc}\ov{\et}_x^ac_x^bc_x^c
\right\}.
\eea
The $\de(t-x_0)$ constraint, appearing because of the time-dependent
variation $\de\la_t$, leads in principle to a non-trivial energy
injection into the ghost lines of the \ST identities (the ghost
functions are, however, discarded in our truncation scheme).  The
above formula can be reexpressed by using the definitions
\eq{eq:deffuncder} for the connected and proper functions, and after
repeating the manipulations from
Ref.~\cite{Watson:2008fb,Popovici:2010mb} one arrives at the following
identity:
\bea
0&=\int d^4x\de(t-x_0)&\left\{
\frac{1}{g}\left(\partial_x^0\ev{\imath\si_x^a}\right)c_x^a
-f^{abc}\ev{\imath\si_x^a}\left[\ev{\imath\ro_x^b\imath\ov{\et}_
x^c}+\si_x^bc_x^c\right]
-\frac{1}{g}\left[\frac{\nabla_{ix}}{(-\nabla_x^2)}\ev{\imath A_
{ix}^a}\right]\ev{\imath\ov{c}_x^a} \right.\nonumber\\
&&
-f^{abc}\ev{\imath A_{ix}^a}t_{ij}(\vec{x})\left[\ev{\imath J_{jx}
^b\imath\ov{\et}_x^c}+A_{jx}^bc_x^c\right]
-\frac{1}{g}\la_x^a\ev{\ov{c}_x^a} +\frac{1}{2}f^{abc}\ev{\imath
c_x^a}\left[\ev{\imath\ov{\et}_x^b
\imath\ov{\et}_x^c}+c_x^bc_x^c\right]
\nonumber\\
&&\left.
+\imath T_{\al\ba}^a\ev{\imath q_{\al x}}\left[\ev{\imath\ov{\chi}
_{\ba x}\imath\ov{\et}_x^a}-c_x^aq_{\ba x}\right] +\imath
T_{\ba\al}^a\left[\ev{\imath\chi_{\ba x}\imath\ov{\et}_x^a}
+c_x^a\ov{q}_{\ba x}\right]\ev{\imath\ov{q}_{\al x}} \right\}.
\eea
Note that functional derivatives involving the Lagrange multiplier
result merely in a trivial identity such that the classical field
$\la_x^a$ can be set to zero \cite{Watson:2008fb}.  Further, one
functional derivative with respect to $\imath c_z^d$ must be taken,
and it follows that the \ST identities are functional derivatives of
(as we will take further derivatives, all the fields and sources must
be retained):
\bea
0&=\int d^4x\de(t-x_0)&\left\{
-\frac{\imath}{g}\left(\partial_x^0\ev{\imath\si_x^d}\right) \de(z-x)
-f^{abc}\ev{\imath\si_x^a}\left[\frac{\de}{\de\imath c_z^d}
\ev{\imath\ro_x^b\imath\ov{\et}_x^c}-\imath\si_x^b\de^{dc}\de(z-x)
\right] \right.\nonumber\\
&& +\frac{1}{g}\left[\frac{\nabla_{ix}}{(-\nabla_x^2)}\ev{\imath
A_{ix}^a}\right]\ev{\imath\ov{c}_x^a\imath c_z^d} -f^{abc}\ev{\imath
A_{ix}^a}t_{ij}(\vec{x})\left[\frac{\de}{\de \imath c_z^d}\ev{\imath
J_{jx}^b\imath\ov{\et}_x^c}-\imath A_{jx}^b \de^{dc}\de(z-x)\right]
\nonumber\\
&& -\imath T_{\al\ba}^a\ev{\imath q_{\al x}}\left[\frac{\de}{\de
\imath c_z^d}\ev{\imath\ov{\chi}_{\ba x}\imath\ov{\et}_x^a}+
\de^{da}\de(z-x)\imath q_{\ba x}\right] \nonumber\\
&&\left.  +\imath T_{\ba\al}^a\left[\frac{\de}{\de\imath
c_z^d}\ev{\imath \chi_{\ba
x}\imath\ov{\et}_x^a}-\de^{da}\de(z-x)\imath\ov{q} _{\ba
x}\right]\ev{\imath\ov{q}_{\al x}}\right\}.
\label{eq:stid0}
\eea

Starting with the above equation, the procedure is now to functionally
differentiate with respect to the quark, antiquark and gluon
fields. In principle, the resulting equation contains a large number
of terms, however most of them simplify in the heavy mass limit and
under truncation. To be specific, after taking the functional
derivatives there are four categories of terms entering the equation
--- three of them are vanishing and one remains.  Firstly, the terms
multiplied by a spatial quark-gluon vertex do not contribute, since
this vertex is suppressed at leading order in the mass expansion (as
explained previously in Sec.~\ref{sec:hqme}).  Secondly, the terms
containing a 4-point function $\G_{\bar qqA\si}$ are also of ${\cal
O}(1/m)$, due to the fact that in the corresponding \DS equation at
least one vertex in each loop term must be at tree-level (just as for
the $\G_{\bar qqA}$ vertex), and we are at liberty to chose this to be
$\G_{\bar qqA}^{(0)}$ (suppressed by the mass expansion) or a pure \YM
contribution (explicitly truncated out). Thirdly, the ghost kernels
arising from the functional derivatives also vanish, since they
necessarily involve \YM vertices. Hence, under truncation and in the
heavy mass limit, only the terms that involve a temporal quark-gluon
vertex will survive. Explicitly, the equation \eq{eq:stid0}, from
which the \ST identity for the quark-2 gluon vertex is derived,
reduces to:
\bea
0=\int d^4x\de(t-x_0) \de(z-x) \left\{
-\frac{\imath}{g}\left(\partial_x^0\ev{\imath\si_x^d}\right)
+f^{abd}\ev{\imath\si_x^a}\imath\si_x^b -\imath T_{\al\ba}^d
\ev{\imath q_{\al x}}\imath q_{\ba x} -\imath T_{\ba\al}^d\imath
\ov{q}_{\ba x}\ev{\imath\ov{q}_{\al x}}\right\}.
\label{eq:stid0tr}
\eea
We now functionally differentiate with respect to $\imath q_{\e
y},\imath \bar q_{\rho t}$ and $\imath\si_{w}^{e}$ and arrive at the
following expression, after setting the sources to zero:
\bea
0=\int dx_{0}\de(t-x_0) &\de(z-x)&\left\{ -\frac{\imath}{g}\pd_z^0
\ev{\imath\bar q_{\rho t}\imath q_{\e y}\imath\si_x^d \imath\si_w^e} +
f^{aed}\ev{\imath\bar q_{\rho t}\imath q_{\e y}\imath\si_x^a} \de(x-w)
\right.\nonumber\\
&&\left.+\imath T^d_{\al\e}\ev{\imath\bar q_{\rho t}\imath q_{\al
x}\imath\si_w^e} \de(x-y) -\imath T^d_{\rho\al}\ev{\imath\bar q_{\al
x}\imath q_{\e y}\imath\si_w^e} \de(x-t) \right\}.
\label{eq:stid5pt}
\eea
Identifying $\ev{\imath\bar q\imath q\imath\si\imath\si} $ with
$\G_{\bar qq\si\si}$ and Fourier transforming, one obtains (the
momentum dependence of the right hand side is trivial)
\be
\G_{\bar qq\si\si}^{de}\sim f^{aed} \G_{\bar qq\si}^{a} +\G_{\bar
qq\si}^{e}\imath T^d- \imath T^d\G_{\bar qq\si}^{e},
\ee
which under truncation (where $\G_{\bar qq\si}=\G_{\bar qq\si}^{(0)}$)
gives
\be
\G_{\bar qq\si\si}^{de}\sim f^{aed} T^{a} +\imath T^e T^d-\imath
T^dT^{e}=\imath[T^e,T^d]+f^{eda}T^a=0.
\ee

\bibliography{4pctfunction}

\end{document}